\documentclass[11pt]{article}
\usepackage{graphicx}                  
\usepackage{amssymb}
\usepackage{amsmath}
\usepackage{bm}
\numberwithin{equation}{section}
\usepackage{multicol}
\usepackage{array}
\newcolumntype{C}[1]{>{\centering\arraybackslash}m{#1}}
\usepackage{epstopdf}
\usepackage{caption}
\usepackage{subcaption}
\usepackage{mathrsfs}
\usepackage{grffile}
\usepackage{floatrow}
\usepackage[amssymb]{SIunits}
\usepackage{floatflt} 
\usepackage[toc,page]{appendix}
\usepackage[listings,theorems]{tcolorbox}
\usepackage[section]{placeins}
\usepackage{wrapfig}
\usepackage{amsthm}
\usepackage[utf8]{inputenc} 
\usepackage{dsfont}
\usepackage{geometry}
\usepackage{setspace}
\usepackage{physics}
\usepackage{authblk}
\usepackage[round]{natbib}
\usepackage{dsfont}
\usepackage[bookmarks]{hyperref}
\usepackage{titlesec}
\usepackage[font=small,labelfont=bf]{caption}
\DeclareMathOperator\supp{supp}
\titleformat*{\section}{\large\bfseries}
\titleformat*{\subsection}{\normalfont\bfseries}
\titleformat*{\subsubsection}{\normalfont\bfseries}

\newtheorem*{theorem*}{Theorem}

\providecommand{\keywords}[1]
{
  \textbf{\textit{Keywords---}} #1
}

\title{\textbf{How (Not) to Understand Weak Measurements of Velocities}}

\author[1]{Johannes Fankhauser\thanks{\href{mailto:johannes.fankhauser@seh.oc.ac.uk}{\it johannes.fankhauser@seh.ox.ac.uk}}, Patrick Dürr\thanks{\href{mailto:patrick.duerr@oriel.ox.ac.uk}{\it patrick.duerr@oriel.ox.ac.uk}}, \it Faculty of Philosophy, University of Oxford, Woodstock Road, Oxford OX2 6GG, UK.}

\date{}   

\begin{document}
\maketitle

 



\begin{abstract}
To-date, the most elaborated attempt to complete quantum mechanics by the addition of hidden variables is the de Broglie-Bohm (pilot wave) theory (dBBT). It endows particles with definite positions at all times. Their evolution is governed by a deterministic dynamics. By construction, however, the individual particle trajectories generically defy detectability in principle. Of late, this lore might seem to have been called into question in light of so-called weak measurements. Due to their characteristic weak coupling between the measurement device and the system under study, they permit the experimental probing of quantum systems without essentially disturbing them. It's natural therefore to think that weak measurements of velocity in particular offer to actually observe the particle trajectories. If true, such a claim would not only experimentally demonstrate the incompleteness of quantum mechanics: it would provide support of dBBT in its standard form, singling it out from an infinitude of empirically equivalent alternative choices for the particle dynamics. Here we examine this possibility. Our result is deflationary: weak velocity measurements constitute no new arguments, let alone empirical evidence, in favour of standard dBBT; One mustn't na\"ively identify weak and actual positions. Weak velocity measurements admit of a straightforward standard quantum mechanical interpretation, independent of any commitment to particle trajectories and velocities. This is revealed by a careful reconstruction of the physical arguments on which the description of weak velocity measurements rests. It turns out that for weak velocity measurements to be reliable, one must already presuppose dBBT in its standard form: in this sense, they can provide no new argument, empirical or otherwise, for dBBT and its standard guidance equation. 
\end{abstract}


\section{Introduction}

Since its inception, Quantum Mechanics (QM) has faced three major interpretative conundrums (see e.g. \citealt{lewis2016quantum, myrvold2018-sep-qt-issues}). The first is the so-called Measurement Problem (see e.g. \citealt{maudlin1995three}): how are we to make sense of the superpositions of states which the formalism of QM (if assumed to be universally valid) appears to attribute to objects? The second pertains to the interpretation of Heisenberg’s uncertainty relations (see e.g. \citealt{sep-qt-uncertainty}): do they circumscribe an absolute limit of simultaneous knowledge of, say, a particle’s momentum and position? Or does it reflect an \textit{ontological} indeterminacy? Finally, how should one understand entanglement (see e.g. \citealt{ney2013wave}) --- the fact that generically, composite systems appear to defy an unambiguous description of their individual constituent parts? 

These three puzzles culminate in the so-called EPR paradox (see e.g. \citealt[Chapter 3]{redhead1987incompleteness} or \citealt{fine-sep-qt-epr}). Suppose one widely separates the partners of an entangled pair of particles. They can then no longer interact. Hence we may, according to Einstein, Podolsky and Rosen, ``without in any way disturbing the system'' perform (and expect a well-defined outcome of) a position measurement on one partner, and a simultaneous momentum measurement on the other \citep[p.~777]{einstein1935can}. Prima facie, it looks as if thereby we can bypass the uncertainty relations. This raises the question whether QM in its current form is complete: does every element of physical reality have a counterpart in the description of the QM formalism?  

Famously, Einstein thought otherwise (see e.g. \citealt{lehner2014einstein}). He was ``[...] firmly convinced that the essentially statistical character of contemporary quantum theory is solely to be ascribed to the fact that this [theory] operates with an incomplete description of physical systems'' \citep[p.~666]{Schilpp1949}.  
To-date, the most elaborated attempt to thus ``complete'' (cf. \citealt[Section~4]{goldstein-sep-qm-bohm}) QM dates back to \citet{bohm1952suggested, bohms1952suggested2} ---  ``Bohmian Mechanics'' or, in recognition of de Broglie's earlier proposal, ``de Broglie-Bohm theory'' (dBBT).\footnote{There exist two \textit{distinct} variants of dBBT: the ``quantum potential''  school (expounded e.g. by \citealt{bohm2006undivided}, or \citealt{holland1995quantum}), and the ``$1
^{st}$-order formulation'', canonised in the oeuvre of Dürr, Goldstein, Zangh\`i and their collaborators. 

The present paper will only be concerned with the latter; ``dBBT'' will exclusively refer to this variant of dBBT, throughout.} (We’ll stick to the latter term throughout.) 

It supplements the QM formalism by a deterministic, but manifestly non-local dynamics for particles. At all times, they occupy determinate positions, evolving continuously in time. Only the particles' initial exact distribution (and the initial wave function) is unknown. Due to this fact, QM emerges from dBBT in a manner ``approximately analogous [...] to the statistical mechanics within the framework of classical mechanics'' --- as Einstein (ibid) had hoped.

But dBBT isn't free of problems. From its early days on, a principal objection to it\footnote{For subtleties in the early objections to dBBT, related to dBBT's unobservability, we refer to \citet{myrvold2003some}} targets the unobservability of its particle dynamics. By construction, in dBBT the individual particle trajectories seem to be undetectable \textit{in principle}. Only their statistical averages are observable. They coincide with the standard quantum mechanical predictions. Thereby, standard dBBT achieves empirical equivalence with QM.\footnote{Here, we'll set aside possible subtleties, see \citet{arageorgis2017bohmian}.}

Recently, this lore seems to have been called into question in light of a novel type of measurements --- so-called weak measurements \citep{aharonov1988result}. These denote setups in which some observable is measured, without significantly disturbing the state.

Inspired by \citet{wiseman2007grounding}, eminent advocates of standard dBBT seem to have touted such weak measurements as a means of actually observing individual trajectories in standard dBBT (e.g. \citealt[Section~4]{goldstein-sep-qm-bohm}). Moreover, they point to already performed experiments (e.g. \citealt{kocsis2011observing, mahler2016experimental}) that appear to corroborate dBBT's predictions and claim to show the particle trajectories. 

The present paper will critically examine those claims. Should they hold up to scrutiny, they would not only establish the incompleteness of QM. Almost more spectacularly, they would also furnish the remedy: they would vindicate dBBT in its standard form.

Those claims, we contend, are mistaken: weak measurements constitute no new arguments, let alone empirical evidence in favour of dBBT's guidance equation. To show this, we'll carefully reconstruct the physical arguments on which the description of weak measurement rests. dBBT is entirely dispensable for a coherent treatment and interpretation of weak measurements; they receive a natural interpretation within standard QM as observational manifestations of the gradient of the wave function's phase.  For weak velocity measurements to disclose the particles' actual velocities, one must not only presuppose the prior existence of deterministic (and differentiable) trajectories, but also the specific form of standard dBBT's particle dynamics. We contest Dürr et al.'s suggestion of a legitimate sense in which weak velocity measurements allow a genuine measurement of particle trajectories.

We'll proceed as follows. §\ref{dBBT} will revisit de Broglie-Bohm theory --- its basics (§\ref{basics}), and one of its principal challenges, its empirical underdetermination (§\ref{section underdetermination}). In §\ref{dBBTweak}, we'll turn to weak velocity values. §\ref{wisevel} will introduce Wiseman's measurement protocol for so-called weak velocity measurements. We'll subsequently illustrate it in the double-slit experiment (§\ref{doubleslit}). Our main analysis of the significance of weak measurements for velocities in de Broglie-Bohm theory will form the subject of §\ref{weak measurements are not genuine}. We'll first elaborate when actual velocities and weak ones (as ascertained in Wiseman's measurement protocol) coincide (§\ref{secton when do weak and actual velocities coincide}). This will enable a critical evaluation both of Dürr et al.'s claim that weak velocity measurements are in some sense genuine (§\ref{DGZgenuine}), and as well as the idea that they provide \textit{non}-empirical support for standard dBBT (§\ref{grounding}). Our findings will be summarised in §\ref{conclusion}. A mathematical appendix (§\ref{Appendix}) contains a concise review of weak interactions within the von Neumann measurement scheme (§\ref{strong measurements appendix}), as well as of post-selection and the two-vector-formalism (§\ref{weak values appendix}).

\section{De Broglie-Bohm Theory}
\label{dBBT}

\subsection{Basics}
\label{basics}

dBBT is best conceived of as an example of what \citet{popper1967quantum} dubbed a ``quantum theory without observer'' (cf. \citealt{goldstein1998quantum}; \citealt[esp. Section~8]{allori2008common}): it aspires to provide an understanding of quantum phenomena without fundamental recourse to non-objective (i.e. subjective or epistemic) notions. 
Such endeavours grew out of the dissatisfaction with influential presentations of QM, notably by von Neumann, Heisenberg and (common readings of) Bohr (see e.g. \citealt{jammer1974philosophy}; \citealt[Ch.~VIII, IX]{scheibe2006philosophie}; \citealt{cushing1996causal}).  

In its non-relativistic form, dBBT is a theory about (massive, charged, etc.\footnote{For our present purposes, we'll elide subtleties concerning the ascription of such intrinsic properties (cf. \citealt{brown1995bohm, brown1996bovine, brown1996cause}). We'll also set aside Esfeld's (\citeyear{esfeld2014quantum, esfeld2017minimalist}) ``Humeanism without properties'' (the ontology and ideology of which is limited to primitively occupied spacetime points and the spatiotemporal relations).}) particles. At all times, they occupy definite positions. Across time, the particles follow deterministic trajectories. Like a ``pilot wave'', the quantum mechanical wave function guides them along those paths. Assuming a particular initial distribution of the particles, one recovers the empirical content of QM. 

More precisely, for an $N$-particle system, dBBT can be taken to consist of three postulates. (We closely  follow \citet{teufel2009bohmian}, to whom we refer for all details.)

\begin{itemize}
    \item[\textbf{(SEQ)}] The wave function $\Psi\colon\mathbb{R}^{3N}\times\mathbb{R}\to \mathbb{C}$ satisfies the standard $N$-particle Schrödinger Equation (SEQ) in the position representation:
    \begin{equation}
        i\hbar \frac{\partial}{\partial t} \Psi(\bm{Q},t)=\hat{H}\Psi(\bm{Q},t) 
    \end{equation} with the $N$-particle Hamiltonian $\hat{H}=-\sum\limits_{i=1}^{N}\frac{\hbar^2}{2m_i}\nabla_i^2 + V(Q,t)$, where $\nabla_i=\frac{\partial}{\partial \bm{Q_i}}$, $i=1,...,N$ acts on the $i$-th position variable $\bm{Q_i}$ and $\bm{Q}:=(\bm{Q_1}, ..., \bm{Q_N})$.
    
    \item[\textbf{(GEQ)}] The continuous evolution of the $i$-th particle's position $\bm{Q_i}(t)\colon \mathbb{R}\to\mathbb{R}^3$ in $3$-dimensional Euclidean space is generated by the flow of the velocity field\footnote{For motivations, see \citealt[Chapter 4]{passon2004bohmsche}.}
    \begin{equation}
    \label{standardguidancelaw}
        v_i^{\Psi}:= \frac{\hbar}{m_i}\Im \frac{\nabla_i \Psi}{\Psi}|_{(\bm{Q_1}(t), ..., \bm{Q_N}(t))}.
    \end{equation}
    That is, the particle position $\bm{Q_i}$ obeys the so-called guidance equation (GEQ)
    \begin{equation}
    \label{guidance equation}
        \bm{\dot{Q}_i}=v_i^{\Psi}.
    \end{equation}
    For all relevant types of potentials, unique solutions (up to sets of initial conditions of measure zero) have been shown to exist \citep{teufel2005simple}. 
    Notice that $v_i^{\Psi}$ depends on all particle positions simultaneously. This is the source of dBBT's manifest action-at-a-distance in the form of an instantaneous non-locality (see e.g. \citealt[Section~13]{goldstein-sep-qm-bohm}).
    \item[\textbf{(QEH)}] The wave function induces a natural (and, under suitable assumptions, unique, see \citet{goldstein2007uniqueness}) measure on configuration space, the so-called Born measure:
    \begin{equation}
        \mathbb{P}^{\Psi}(d^{3N}\bm{Q}):=|\Psi|^2d^{3N}\bm{Q}.
    \end{equation}
    It quantifies which (measurable) sets of particle configurations $\mathcal{Q}\subseteq \mathbb{R}^{3N}$ count as large (``typical''). That is:
    \begin{equation}
        \int_{\mathcal{Q}}d^{3N}\bm{Q}|\Psi(\bm{Q})|^2=1-\varepsilon, \nonumber
    \end{equation} for some small $\varepsilon>0$ (see \citealt{maudlin2011three, durr2019typicality, dustin2015typicality} for details; cf. \citealt{frigg2009typicality, frigg2011typicality}).\footnote{Typicality raises intriguing questions about whether appeal to it is explanatory (and if so, in which sense). For a recent account, see \citet{wilhelm2019typical}.} This definition of typicality respects a generalised sense of time-independence. A universe typical in this sense is said to be in quantum equilibrium (see \citealt{durr1992quantum} for further details). The continuity equation for $|\Psi|^2$ obtained from the Schrödinger Equation implies that a system is in quantum equilibrium at all times, if and only if in equilibrium at \textit{some} point in time. This is called the Quantum Equilibrium Hypothesis (QEH). 

\end{itemize}

Consider now a de Broglie-Bohmian $N$-particle universe, satisfying these three axioms. An $M$-particle subsystem is said to possess an ``effective'' wave function $\Phi$, if the universal wave function (i.e. the wave function of the universe) $\Psi\colon X\times Y \to \mathbb{C}$, with $X$ and $Y$ denoting the configuration space of the subsystem and its environment, respectively, can be decomposed as

\begin{equation}
    \forall (x,y) \in X\times Y\colon \Psi(x,y) = \psi(x)\Phi(y) + \Psi_{\perp}(x,y).\nonumber
\end{equation} Here $\Phi$ and $\Psi_{\perp}$ have macroscopically disjoint $y$-support and $Y \subseteq \supp (\Phi)$.  That is, the configurations in which $\Phi$ and $\Psi_{\perp}$ vanish are macroscopically distinct (e.g. correspond to distinct pointer positions). For negligible interaction with their environment, the effective wave function $\psi$ of subsystems can be shown to satisfy the Schrödinger Equation itself.

\subsection{Underdetermination}
\label{section underdetermination}
Empirically, the guidance equation \ref{guidance equation} isn't the only option.

More precisely, for empirical equivalence with QM, the specific guidance equation \ref{guidance equation} isn't necessary. Infinitely many different choices 
\begin{equation}
\label{altguidanceequation}
    \bm{v}^{\Psi}\mapsto \bm{v}^{\Psi}+ |\Psi|^{-2}\bm{j}
\end{equation} are equally possible for otherwise arbitrary vector fields $\bm{j}$ whose divergence vanishes, $\nabla \cdot \bm{j}=0$. They yield coherent alternative dynamics with distinct particle trajectories, whilst leaving the predictive-statistical content unaltered \citep{deotto1998bohmian}. 

One needn't even restrict oneself to a deterministic dynamics (an option expressly countenanced by e.g. \citealt[Chapter~1.2]{teufel2009bohmian}): a stochastic dynamics, with $|\Psi|^{-2}\bm{j}$ corresponding to a suitable random variable can also be introduced. As a result the particles would perform random walks, with the r.h.s. of the integral equation
\begin{equation}
\nonumber
\bm{Q}(t)-\bm{Q}(t_0)=\int\limits_{t_0}^{t}\bm{v}^{\Psi}d\tau
\end{equation} containing a diffusion term. A proposal of this type is Nelson Stochastics (see e.g. \citealt{goldstein1987stochastic}; \citealt{Bacciagaluppi2005intronelson}). In short: by construction, dBBT's individual particle trajectories are observationally inaccessible.

In consequence, dBBT is vastly underdetermined by empirical data: all versions of dBBT with guidance equations of the type \ref{altguidanceequation} are experimentally indistinguishable. Yet, the worlds described by them clearly differ. (We illustrate this in Figure \ref{fig:trac}.) 

\newsavebox{\smlmat}
\savebox{\smlmat}{$\bm{j}:=\frac{1}{x^2+y^2}\left(\begin{array}{c} 
	-y\\
	x 
	\end{array}\right)$}

\begin{figure}[ht]
	\centering
	\begin{subfigure}[b]{0.4\textwidth}
		\includegraphics[width=\textwidth]{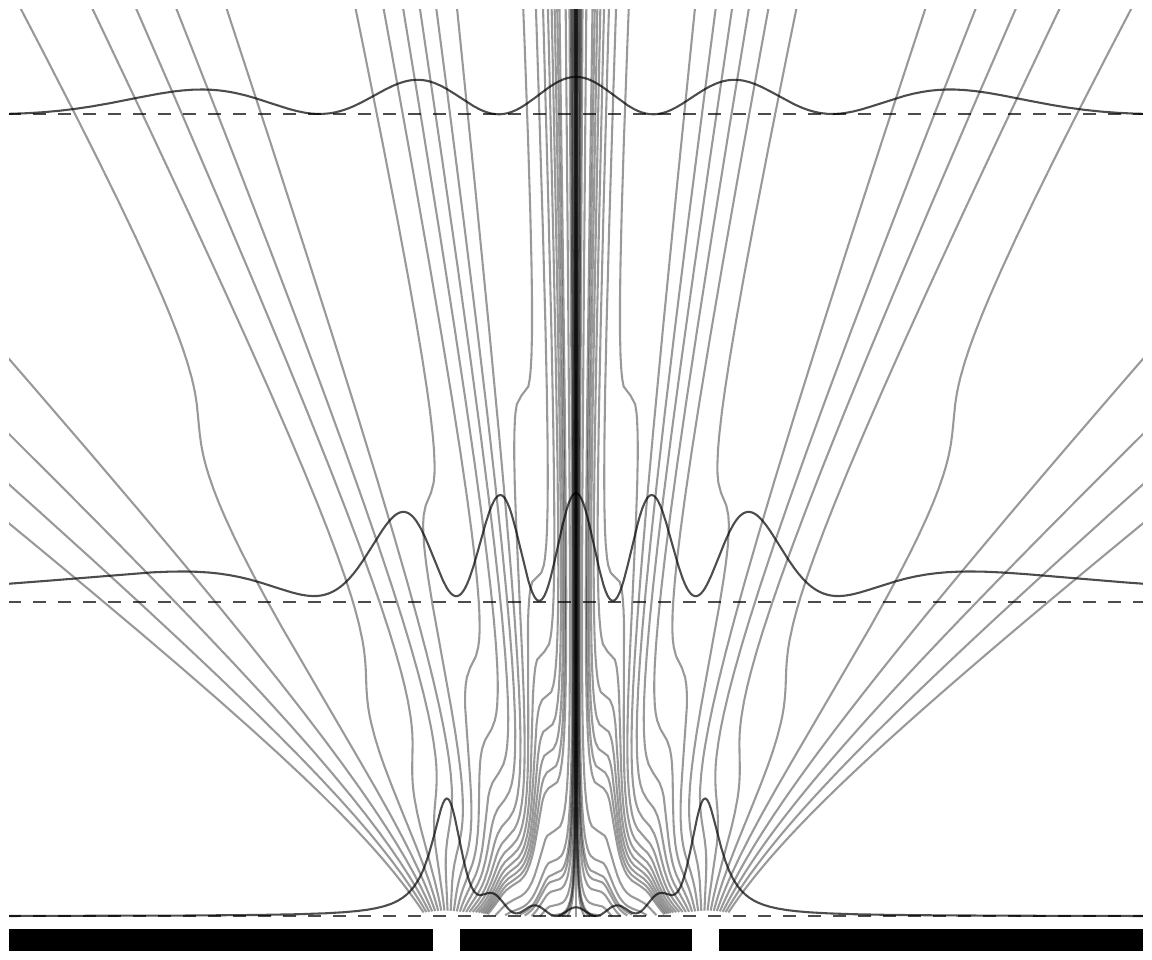}
		\caption{}
		\label{fig:s1}
	\end{subfigure}
	~
	\begin{subfigure}[b]{0.4\textwidth}
		\includegraphics[width=\textwidth]{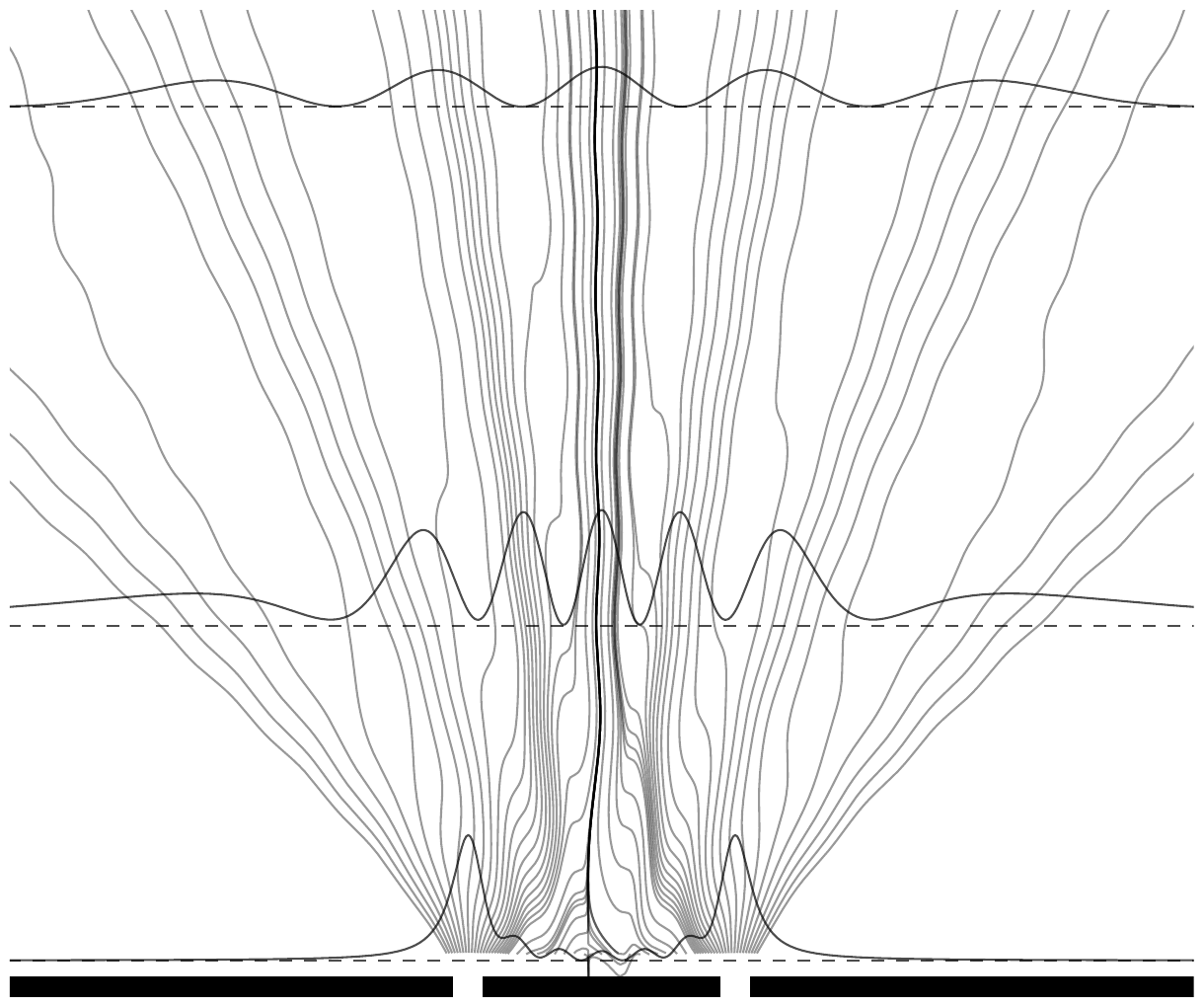}
		\caption{}
		\label{fig:s2}
	\end{subfigure}
	~
	\begin{subfigure}[b]{0.4\textwidth}
		\includegraphics[width=\textwidth]{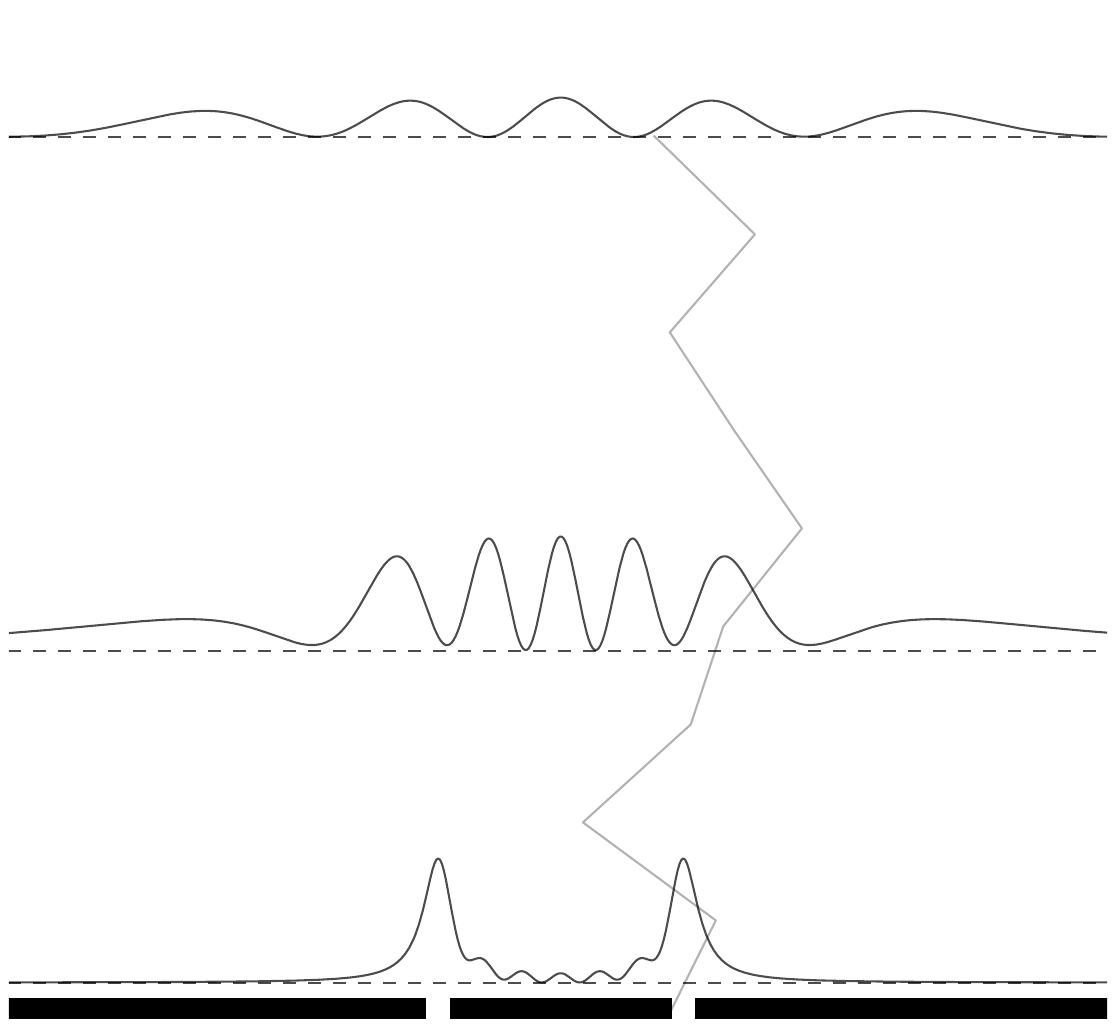}
		\caption{}
		\label{fig:s3}
	\end{subfigure}
	
	\caption{A particle follows different trajectories corresponding to different/non-standard guidance equations.
	\textbf{(a)} The familiar wiggly deterministic trajectories that lead to the interference pattern in a double-slit experiment determined by the standard guidance equation. 
	\textbf{(b)} Alternative trajectories obtained from adding a divergence-free vector field~\usebox{\smlmat} to the standard Bohmian velocity field.  
	\textbf{(c)} A single stochastic trajectory generated by a random variable sampled according to $|\psi|^2$. For illustration, the probability density $|\psi|^2$ is shown at three different snapshots in time.
	All  choices of the dynamics (i.e. (a), (b), (c)) are observationally indiscernible: The resulting measurable distributions at the screen at the top of each figure are the same.}
	\label{fig:trac}	
\end{figure} 

This underdetermination poses a challenge to a realist understanding of dBBT (cf. for example \citealt[Chapter~3.2]{Stanford-sep-scientific-underdetermination})

For the purposes of this paper, we'll confine the class of considered choices to the family of de Broglie-Bohmian-like theories (cf. \citealt[Section 3.4]{durr2017probabilities}) --- i.e. particle theories within the Primitive Ontology framework (see, e.g. \citealt{allori2008common,allori2013primitive,allori2015primitive}). It encompasses e.g. the ``identity-based Bohmian Mechanics'' \citep{goldstein2005all} or ``Newtonian QM'' \citep{sebens2015quantum}. Let's even further whittle down the list of candidate theories to deterministic variants of dBBT with differentiable paths, i.e. to variants of dBBT  that differ only with respect to their vector field of the type in Equation \ref{altguidanceequation}. Still, the underdetermination persists; its severity is scarcely diminished: how to justify the particular choice for the standard guidance equation amongst the uncountably infinite alternatives?

An argument frequently cited in response is a result by \citet[p.~852]{durr1992quantum}: ``The standard guidance equation is the simplest first-order equation that respects Galilei covariance and time-reversal invariance.''

But this is not decisive. First, individually neither desideratum of Dürr et al.'s theorem seems compelling --- unless one is already guided by intuitions, shaped by either classical physics or by standard dBBT itself. In particular, one may reject the ab initio requirement of Galilei covariance as implausible: Galilei covariance is the symmetry group of \textit{Classical} Mechanics.\footnote{Not even this is entirely obvious. On the one hand, at least once one incorporates Newtonian gravity, the most perspicuous spacetime setting is no longer Galilei spacetime (e.g. \citealt{pooley_subst_2013}). On the other hand, one may be attracted to the idea of a theory of classical mechanics that incorporates the Leibniz Group as its symmetry group, such as in the Barbour-Bertotti theory (ibid., Section 6.2). (Notice that recently attempts have indeed been made, e.g. by \citet{vassallo2015can}, to combine Barbourian shape dynamics with dBBT.)

In either case, the symmetry groups would be larger than the Galilei group.}

Why impose it on a more fundamental theory --- dBBT --- which is supposed to \textit{supersede} Classical Mechanics?\footnote{From the perspective of the so-called dynamical approach (e.g. \citealt{brown2005physical, brown2018dynamical}) to spacetime symmetries, this requirement lacks apriori force. According to advocates of this approach, spacetime symmetries merely codify (are reducible to) the symmetries of the matter dynamics. For them, therefore, to demand any particular spacetime symmetry as an ab initio constraint on a possible, fundamental matter dynamics is to put the cart before the horse (cf. \citealt{acuna2016minkowski, myrvold2017could}).}

Secondly, Dürr et al.'s argument rests on an assumption about how the Galilei group acts on the wave function. As \citet{skow2010symmetry} has argued, such an assumption is essentially unwarranted.

Thirdly, let's grant that a satisfactory answer can be given to the preceding two questions. Dürr et al.'s argument pivotally turns on mathematical simplicity. We confess, we'd be hard-pressed to pinpoint what \textit{mathematical} (rather than, say, ontological) simplicity precisely means. Whatever it may be, as a super-empirical criterion, it may well be felt a dubious indicator of truth (see e.g. \citealt[Chapter~4.4]{van1980scientific}; \citealt{norton2000nature}, \citealt[Chapter~5-7]{norton2018material}; \citealt{ivanova2014there,ivanova_forthcoming}). At best we are inclined to regard it as a pragmatic criterion, at worst an aesthetic one for theory acceptance. Can a realist legitimately invoke it to argue that one theory is more likely to be true than an otherwise equally good alternative? 

This context --- underdetermination --- renders weak value measurements particularly interesting. By (prima facie) allowing measurements of individual particle trajectories, they appear to directly overcome dBBT's underdetermination. But wouldn't that contradict the empirical inaccessibility of the trajectories? Let us see.

\section{Weak velocity values}
\label{dBBTweak}

This section will offer a concise review of so-called weak values. We'll first outline how they are harnessed in Wiseman's measurement protocol for weak velocity measurements (§\ref{wisevel}). An application to the double-slit experiment will further illustrate the salient points (§\ref{doubleslit}). This will pave the way for our subsequent discussion in
(§\ref{weak measurements are not genuine}).

\subsection{Wiseman's measurement protocol for weak velocity measurements}
\label{wisevel}

Following \citealt{aharonov1988result}, weak measurements are measurement processes (modelled via the von Neumann scheme, see §\ref{strong measurements appendix}) in which the interaction between the measurement apparatus (``pointer device'') and the particle (``system'') is weak: it disturbs the wave function only slightly. As a result, one can extract still further information about the particles (say, regarding their initial momenta) via a subsequent ordinary ``strong'' (or projective) measurement (say, regarding their positions).

More precisely: after a weak interaction (say, at $t=0$), the pointer states aren't unambiguously correlated with eigenstates of the system under investigation. In contradistinction to strong measurements, the system doesn't (effectively) ``collapse'' onto eigenstates; the particles can’t be (say) located very precisely in a single run of an experiment. This apparent shortcoming is compensated for when combined with a strong measurement a tiny bit \textit{after} the weak interaction: the experimenter is then able not only to ascertain the individual particle’s precise location (via the strong measurement); for a sufficiently large ensemble of identically prepared particles with initial state $\psi_{in}$ (viz. Gaussian wave packets with a large spread), she can also gain statistical access to the probability amplitude of all subensembles whose final states --- the so-called ``post-selected'' state --- have been detected (in the strong measurement) to be $\psi_{fin}$:

\begin{equation}
\langle \hat{x}\rangle _ w:=\frac{\langle \psi_{fin}|\hat{x}|\psi_{in}\rangle}{\langle\psi_{fin}|\psi_{fin}\rangle}
\end{equation}

This quantity is called the ``weak position value'' (for the position operator $\hat{x}$). (The concept is straightforwardly applied also to other operators, mutatis mutandis.) It can be shown (see \ref{weak values appendix}) that after many runs, the pointer’s average position will have shifted by $\langle \hat{x} \rangle _w$.
Specifically, if we characterise the final/post-selected state via position eigenstates $|x\rangle$, determined in a strong position measurement and unitary evolution of the initial state, we obtain 

\begin{equation}
\langle \hat{x} (\tau)\rangle _ w=\Re\left(\frac{\langle x|\hat{U}(\tau)\hat{x}|\psi_{in} \rangle}{\langle x|\hat{U}(\tau)|\psi_{in} \rangle}\right),
\end{equation}

where $\hat{U}(\tau)$ denotes the unitary time evolution operator during the time interval $[0;\tau]$. Following \citealt{wiseman2007grounding}, it’s suggestive to construe $\langle \hat{x}(\tau) \rangle _w$ as the mean displacement of particles whose position was found (in a strong position measurement at $t=\tau$) to be at $x$. From this displacement, a natural definition of a velocity field ensues:

\begin{equation}
\label{operational velocity}
\textbf{v}(\textbf{x},t)= \lim\limits_{\tau \rightarrow 0}\frac{1}{\tau}(\textbf{x}-\langle\hat{x}_w\rangle).
\end{equation} 

Note that all three quantities entering this velocity field --- $\tau$, $x$ and $\langle \hat{x} (\tau)\rangle _w$ --- are experimentally accessible. In this sense, the velocity field is ``defined operationally'' (Wiseman). In what follows, we’ll refer to the application of this measurement scheme --- a strong position measurement in short succession upon a particle’s weak interaction with the pointer --- for the associated ``operationally defined'' velocity field  as ``Wiseman’s measurement protocol for weak velocity measurements'', or simply \textit{``weak velocity measurements'}'.

For a better grasp of its salient points, let’s now spell out such weak velocity measurements in the context of the double-slit experiment. In §\ref{weak measurements are not genuine}, it will prove useful to occasionally refer back to this concrete setup.

\subsection{Weak measurements in the double-slit experiment}
\label{doubleslit} 

Consider the standard double-slit experiment with, say, electrons, hitting a screen. It enables a detection of the electrons' positions. This constitutes a strong position measurement. Accordingly, we'll dub this screen the \textit{strong screen}. Between the strong screen and the two slits from which the particles emerge, let a weak measurement of position be performed. Let this be called the \textit{weak screen}. The two screens can be moved to perform measurements at various distances from the double-slit. Suppose that it takes the particles some time $\tau>0$ to travel from the weak to the strong screen. 

After passing through the slits, the electron will be described by the wave function $\ket{\psi}=\int \psi(x,t)\ket{x}dx$. This leads to the familiar double-slit interference fringes. We assume that the weak screen, i.e. the pointer variable, is in a Gaussian ready state with width $\sigma$, peaked around some initial position. After the particles have interacted with the measurement device (at time $t=0$), the composite wave function $\ket{\Psi(0)}$ of particle-\textit{cum}-weak screen is 

\begin{equation}
\label{state}
\ket{\Psi(0)}=\int \psi(x,t)\ket{x}\otimes \varphi(y-x)\ket{y}dxdy.
\end{equation} 

Here, $\ket{\varphi}$ denotes the wave function of the weak screen, and $y$ its free variable  (e.g. the position of some pointer device). The wave function then evolves unitarily for some time $\tau$, according to the particle Hamiltonian $\hat{H}$: 

\begin{equation}
\ket{\Psi(\tau)}=\hat{U}(\tau)\ket{\Psi(t)}=e^{-\frac{i}{\hbar}\hat{H}\tau}\ket{\Psi(0)}.
\end{equation} 

After weakly interacting, the particle and pointer are entangled. Hence, only the composite wave function --- \textit{not} the reduced state of the pointer --- evolves unitarily during time $\tau$. The unitary operator $\hat{U}(\tau):e^{-\frac{i}{\hbar}\hat{H}\tau}$ only acts on $x$ (not on $y$). Due to this evolution, the post-selected position $\textbf{x}$ on the strong screen will in general differ from the weak value $\langle\hat{x}_w\rangle$, obtained from averaging the conditional distribution of the pointer of the weak screen. The procedure is depicted in Figure \ref{fig:correspondence}. 

On both screens the wave function is slightly washed out. It evidently differs from an undisturbed state (i.e. in the absence of the weak screen). To obtain the two  position values --- the weak and the strong one --- strong measurements are now performed both at the weak and the strong screen (i.e. on the pointer variable and on the target system). For each position outcome $x$ at the strong screen, let's select a subensemble. For any such subensemble, we then read out the statistical distribution of the position measurement outcomes at the weak screen. 

We have thus assembled all three observable quantities needed for Wiseman's operationally defined velocity \ref{wisevel}: the time $\tau$ that elapsed between the two measurements, the positions $x$ (obtained as values at the strong screen), and the average value of all positions of the subensemble, associated with (i.e. post-selected for) $x$.  
This may now be done for different positions $x$ on the strong screen. To that end, move the screens to different locations; there repeat the measurements. 

With this method one can eventually map the velocity field, for a sufficiently large number of measurements. We'll defer the discussion of how to construe this result to the next section. For now, let's rest content with stating it as a calculational fact, suspending any further conclusions. 

\begin{figure}[H]
	\centering
	\includegraphics[width=0.9\textwidth]{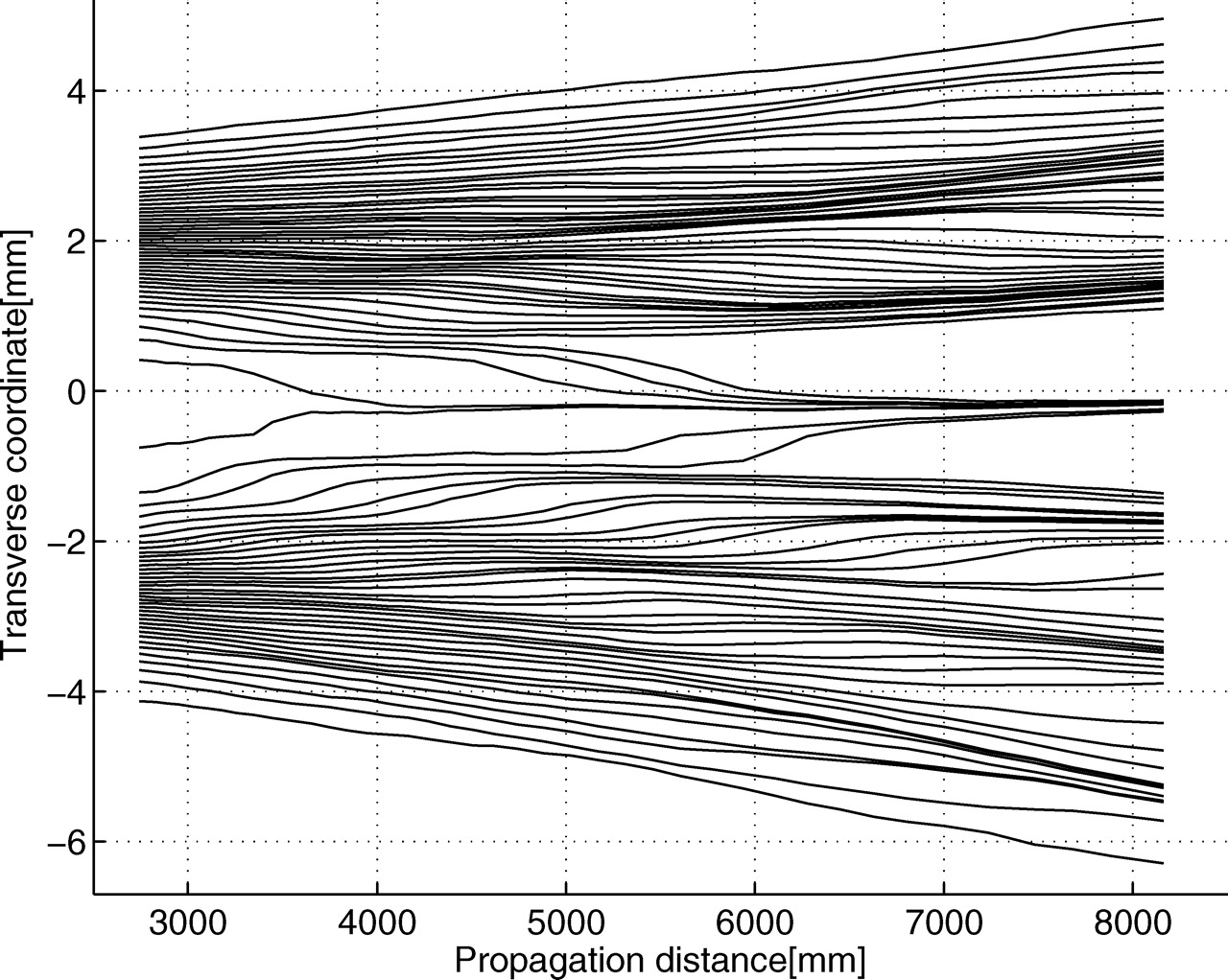}
	\caption{A weak velocity measurement for photons allows the reconstructions of trajectories, qualitatively identical to those of particles in standard dBBT. Particle trajectories in a double-slit experiment performed by \citealt{kocsis2011observing}.}
	\label{fig:Kocsis}
\end{figure}

Kocsis et al. have indeed performed an experiment of a similar kind, using  weak measurements of momentum. Their result, depicted in Figure \ref{fig:Kocsis}, qualitatively reproduces the trajectories of standard dBBT. (We'll return to this experiment and how to understand it in §\ref{weak measurements are not genuine}. Here, we mention it primarily to convey an impression of the qualitative features of Wiseman's operational velocity, when experimentally realised.) Moreover, it can be shown (cf. §\ref{weak velocity and the gradient of the phase}) that weak velocity measurements are measurements of the gradient of the phase of the wave function. Thus, they coincide with definition of standard Bohmian velocities in the guidance equation

\begin{equation}
    v= \frac{\hbar}{m}\nabla S, 
\end{equation} where $S$ is the gradient of the phase of the wave function, $\psi(x)=|\psi|e^{i S(x)}$. 

Notice that for this, only the standard quantum-mechanical formalism has been utilised. Therefore, we may conclude that --- based solely on standard QM --- weak velocity measurements permit experimental access to the gradient of the wave function's phase. 

Next, we'll ponder whether commitment to \textit{further}, generic and supposedly mild interpretative choices (viz. the adoption of a de Broglie-Bohmian framework) might grant us a peep into an allegedly deeper reality, veiled under this standard quantum mechanical interpretation.

\section{Why weak velocity measurements \textit{do not} measure velocities}
\label{weak measurements are not genuine}
  
Suggestive as these results are, we will now show that such measurements could not provide direct experimental evidence displaying the shape of particle trajectories, even if it is assumed that some deterministic particle trajectories exist. They cannot, that is, go any way to experimentally resolving the underdetermination in putative dBBT guidance equations mentioned previously. First (§\ref{secton when do weak and actual velocities coincide}), we'll analyse the relation between Wiseman's operationally defined velocity Equation \ref{operational velocity} and the particle's actual velocity. In particular, we'll show that a strong assumption is required that would render it question-begging to employ weak velocity measurements in order to infer the particles' actual velocities. This analysis will subsequently allow us to critically evaluate two stances regarding the significance of weak velocity values for dBBT --- Dürr et al.'s portrayal of weak velocity measurements as allegedly ``genuine'' measurements (§\ref{DGZgenuine}), and a view of weak velocity measurements as \textit{non}-empirical support of standard dBBT (§\ref{grounding}).

\subsection{When do weak and actual velocities coincide?}
\label{secton when do weak and actual velocities coincide}
Here, we'll address the question of whether --- or rather: \textit{when} --- weak velocities coincide with the particles' actual velocities, assuming that they exist. That is, we'll explicate the conditions under which weak velocity measurements count as reliable. That, we'll argue, turns out to \textit{presuppose} standard dBBT.  

In the following, $x$ and $y$ will denote the position variables of the individual particles to be measured, and the measurement apparatus, respectively. For simplicity, we'll work in one dimension only. Let the particles be prepared in the initial state 
\begin{equation}
   \ket{\psi}=\int dx ~\psi(x)\ket{x}.
\end{equation}  

Furthermore, let the pointer device (i.e. the  weak screen of the double-slit version of weak measurements in §\ref{doubleslit}) be prepared in the initial state given by a Gaussian with large spread $\sigma$, centred around $0$: 
\begin{equation}
    \ket{\varphi}=\int dy ~\varphi(y)\ket{y}=N\int dy e^{-\frac{y^2}{4\sigma^2}}\ket{y},
\end{equation} where $N$ is a suitable normalization factor. Together, the particle and the pointer form the compound system with the joint initial state

\begin{equation}
    \ket{\psi}\otimes\ket{\varphi}= \int dxdy ~\psi(x)\varphi(y)\ket{x}\otimes\ket{y}.
\end{equation}

Now consider the first --- the weak --- measurement process. It consists of an interaction between the particle and the pointer. Upon completion of this process (say at $t=0$), the compound system ends up in the entangled state

\begin{equation}
    \ket{\Psi(x,y,0)}= \int dxdy ~\psi(x)\varphi(y-x)\ket{x}\otimes\ket{y}.
\end{equation}

The probability distribution for the pointer variable $y$, \textit{given} some position $X$ of the particle, is therefore:

\begin{equation}
    \rho_X(y)=\frac{|\Psi_0(X,y)|^2}{|\psi(X)|^2}=|\varphi(y-X)|^2.
\end{equation} This probability density determines the expectation value 

\begin{equation}
\label{expectation value at t=0}
\mathbb{E}(y|x=X)=\int dy~y \rho_X(y)=X.    
\end{equation} That is, the mean value of the pointer distribution, conditional on the particle occupying position $X$, coincides with that position. This underwrites the following counterfactual:

\begin{itemize}
\label{C_0}
    \item[$\mathbf{(C_0)}$] \textit{If one were to perform an ordinary (strong) position measurement on the particles \textit{immediately after} the weak interaction, the expectation value would yield the actual position of the particle.}
\end{itemize} 

Via $\mathbb{E}(y|x=X)$, the particle position thus is empirically accessible through the statistics of large ensembles of identically prepared particles from which we cull post-selected outcomes $x=X$.  

This thought is further exploited in the final steps of Wiseman's procedure. In the foregoing considerations, the strong measurement was performed immediately upon the weak one. Instead, we'll now allow for a small delay. That is, after the particle and the pointer have (weakly) interacted, the total system evolves freely for some small, but finite time $\tau$. Its state then is

\begin{equation}
    \ket{\Psi(x,y,\tau)}= e^{-\frac{i}{\hbar}\tau\hat{H}_0}\ket{\Psi(x,y,0)},
\end{equation} where $\hat{H}_0$ denotes the system’s free Hamiltonian. 

Eventually, we perform a strong measurement of the particle’s position $X_{\tau}$ at $t=\tau$. (The strong coupling between the measurement device and the particle enables a precise detection of the latter's actual position.) We thus get the expectation value for the pointer variable, conditional on the particle occupying the position $X_{\tau}$ at $t=\tau$:

\begin{equation}
\label{weak position value}
    \mathbb{E}(y|x=X_{\tau})=\int dy~y |\Psi(X_{\tau},y,\tau)|^2.
\end{equation}

Through the statistics of a sub-ensemble of particles whose strong position measurements yielded $X_{\tau}$, this expectation value is empirically accessible.

In \textit{analogy} to Equation \ref{expectation value at t=0}, let's define the position: 

\begin{equation}
\label{position}
    X_0:=\mathbb{E}(y|x=X_{\tau}).
\end{equation}

Combined with the particle position $X_\tau$, obtained from the strong measurement at $t=\tau$, it thus appears as if we have access to particle positions at two successive moments. Using Equation \ref{position}, the associated displacement is

\begin{equation}
\label{displacement}
    X_{\tau}-X_0=X_{\tau}-\mathbb{E}(y|x=X_{\tau})
\end{equation}

Let's grant one can make it plausible that the particles' trajectories are differentiable. Then, the displacement (Equation \ref{displacement}) gives rise to the velocity field 

\begin{equation}
\label{velocity0}
v(X_0):= \lim\limits_{\tau \rightarrow 0}\frac{1}{\tau} (X_{\tau} - \mathbb{E}(y|x=X_{\tau})).
\end{equation}

Note that all terms on the r.h.s. of Equation \ref{velocity0} are observable. (Hence, presumably, Wiseman's labelling \ref{velocity0} as an ``operational definition''.) In conclusion, it seems, via the statistics of an experimental setup implementing Wiseman’s procedure, we are able to empirically probe this velocity field. 

But what does this velocity field signify? It's tempting to identify it with the particles' actual velocities. That is, should this be true, the flow of Equation \ref{velocity0} generates the particles' trajectories (assumed to be deterministic and differentiable). Is this identification justified?

By \textit{defining} an $X_0 \overset{def}{=}\mathbb{E}(y|x=X_{\tau})$ via Equation \ref{position}, our notation certainly suggests so. Let’s indeed assume that this is correct. We’ll dub this the ``Correspondence Assumption'' (COR). That is, suppose that the actual particle position $X_{\tau}$ at $t=\tau$ is connected with the earlier particle position $x(0)=X_0=\hat{T}_{-\tau}X_{\tau}$ at $t=0$,  where $\hat{T}_{-\tau}$ denotes the shift operator that backwards-evolves particle positions by $\tau$. (In other words: for arbitrary initial positions, $\hat{T}_{\tau}$ supplies the full trajectory.) Then, according to (COR), the expectation value (\ref{position}) corresponds to the particles' position at $t=0$. For post-selection of subensembles with $x(\tau)=X_{\tau}$ , (COR) thus takes the form (in the limit of large spread $\sigma$):

\begin{equation}
    \textbf{\text{(COR)}}~  \mathbb{E}(y|x(\tau)=X_{\tau})=\hat{T}_{-\tau}X_{\tau}.
\end{equation}

In other words, (COR) implies the counterfactual:

\begin{itemize}
\label{C_t}
    \item[$\mathbf{(C_t)}$] \textit{If one were to perform a strong position measurement at $t=\tau$ (with the weak interaction taking place at $t=0$), yielding the particles' position at $x(\tau)=X_{\tau}$, the weak value would be directly correlated with the particles' earlier position $\hat{T}_{-\tau}X_{\tau}$. That is, upon a strong measurement at $t=\tau$, the expectation value would reveal the particles' true positions:
    \begin{equation}
        \mathbb{E}(y|x(\tau)=X_{\tau})=\hat{T}_{-\tau}X_{\tau}.
    \end{equation}}
\end{itemize} 

On (COR), the weak value thus gives the particle’s \textit{actual} position at the weak screen: the expectation value on the l.h.s. is reliably correlated with the particle's earlier positions. But most importantly, this is an \textit{if and only if condition}: If (COR) is satisfied, then we recover the actual position, but if it is not, we don't. As a result one ought to have to assume that (COR) is true for weak position measurements to yield actual particle positions. 

Thereby, any set of data compatible with QM appears to corroborate standard dBBT: \textit{given (COR)}, weak velocity measurements yield standard dBBT's velocity field. It thus seems as if standard dBBT’s empirical underdetermination has been overcome.

Such an apparent possibility of confirming standard dBBT would be remarkable. It crucially hinges, however, on the soundness of (COR). Why believe that it’s true? We’ll first refute a prima facie cogent argument for (COR). We’ll then give a more general argument why (COR) is generically false. This will eventually be illustrated with a simple counterexample.
Prima facie, (COR) looks like a plausible extrapolation of a strong measurement immediately after the weak interaction (i.e. at $t=0$). This idea may be developed in three steps. First, (COR) indeed holds in the limit $\tau\rightarrow 0^+$. Next, in a deterministic world, it would seem that

\begin{equation}
    \mathbb{E}(y|x(\tau)=\hat{T}_{\tau}\kappa)=\mathbb{E}(y|x(0)=\kappa),
\end{equation} where $\kappa\in\mathbb{R}$ denotes a position. 

By appeal to $C_0$, this would then yield 

\begin{equation}
    \mathbb{E}(y|x(\tau)=\hat{T}_{\tau}\kappa)=\mathbb{E}(y|x(0)=\kappa)=\kappa,
\end{equation} as desired.

At first blush, this argument looks watertight. Its first step ensues from the standard rules of QM (see Equation \ref{expectation value at t=0}). Its third step, too, seems innocuous: only a few lines earlier, we derived $(C_0)$ from the standard QM formalism. Let’s therefore throw a closer glance at the second step. It’s convenient to cast it in terms of the probability densities, associated with the expectation values:

\begin{equation}
    \mathbb{P}(y|x(\tau)=\hat{T}_{\tau}\kappa)=\mathbb{P}(y|x(0)=\kappa).
\end{equation}

Prima facie, given determinism, this identity stands to reason: all else being equal, the probability of craving a biscuit around $5$ pm, given our momentary glucose levels, isn't altered by conditioning on our glucose levels a few minutes earlier (provided that they evolve deterministically). Determinism ensures that those physiological states (and \textit{only} they) evolve into the physiological states, considered initially. By the same token, one might think, the events $\{(x(\tau),y)\in \mathbb{R}\times\mathbb{R}:x(\tau)=\hat{T}_{\tau}\kappa\}$ and $\{(x(0),y)\in \mathbb{R}\times\mathbb{R}:x(0)=\kappa\}$ refer to the same events of our probability space (i.e. the same diachronically identical configurations, as it were, merely pointed to via (inessential) different time indices) and \textit{therefore} are assigned the same probability measure.

Yet, this inference is illicit. While it’s true that $\{(x(\tau),y)\in \mathbb{R}\times\mathbb{R}:x(\tau)=\hat{T}_{\tau}\kappa\}$ and $\{(x(0),y)\in \mathbb{R}\times\mathbb{R}:x(0)=\kappa\}$ contains the same pointer configurations, this \textit{doesn't} imply that $\mathbb{P}(y|x(\tau)=\hat{T}_{\tau}\kappa)=\mathbb{P}(y|x(0)=\kappa)$. For this to hold, the conditional probabilities --- as defined via post-selection --- on both sides must be well-defined. That is, 
\begin{equation}
\label{conditional probabilities}
    \frac{\mathbb{P}(y \& x(\tau)=\hat{T}_{\tau}\kappa)}{\mathbb{P}(x(\tau)=\hat{T}_{\tau}\kappa)} \ \text{and} \ \frac{\mathbb{P}(y \& x(0)=\kappa)}{\mathbb{P}(x(0)=\kappa)}
\end{equation} must exist (and coincide). 

In classical Statistical Mechanics, one may take this for granted. In a \textit{quantum} context, entanglement complicates the situation: it compromises the ascription of probability measures to certain events. One must heed the time with respect to which the assigned probability measure is defined. This is the case with weak velocity measurements. Recall that in Wiseman’s measurement protocol, the strong measurement is only performed at $t=\tau$. This precludes defining the second term in \ref{conditional probabilities}! That is, no strong measurement is performed --- and no attendant ``effective collapse'' of the wave function occurs --- at an \textit{earlier} time (viz. at $t=0$). As a result, at the time of the weak interaction ($t=0$), the wave function of the pointer and that of particles are entangled. That means, however, that we \textit{can't} na\"ively assign the event of any particular particle position at $t=0$ an objective, individual probability measure\footnote{In this regard, one should bear in mind that, on the mainstream view of dBBT (espoused by DGZ), probabilistic statements about subsystems (construed in terms of typicality) should be \textit{derived} from dBBT‘s axioms of §\ref{dBBT} (cf. \citealt[Chapter~4, 9]{teufel2005simple}; \citealt{dustin2015typicality}; \citealt[Section~3]{lazarovici2018observables}). 

In the present context in particular, we can’t simply assign the particles a probability measure --- that of the reduced density matrix --- \textit{per stipulation}: we must deduce it from the probability measure of the composite pointer-device system --- using only the other axioms. The quantum operation of a partial trace, implementing the transition to the reduced density matrix, transcends those fundamental axioms (see e.g. \citet[Section~6]{Durr2004}).}; that would require post-selection at $t=0$. Only the entangled pointer-\textit{cum}-particle system as whole has a physically grounded, objective probability measure.

This follows from the fact that $\mathbb{P}(x(0)=\kappa)$ is obtained from the pointer-\textit{cum}-particle system’s reduced density matrix (i.e. by partially tracing out the pointer’s degrees of freedom). But this transition from the density matrix of a pure state to the reduced density matrix of an ``improper mixture'' \citep[Chapter~7]{espagnat2018conceptual} lacks objective-physical justification (see, e.g., \citealt[Chapter~3-4]{mittelstaedt2004interpretation}). Contrast that with the situation of $\frac{\mathbb{P}(y \& x(\tau)=X_{\tau})}{\mathbb{P}(x(\tau)=X_{\tau})}$: this \textit{is} well-defined via post-selection. That is, due to the ``effective collapse'' (see, e.g., \citealt[Chapter 9.2]{teufel2009bohmian}), induced by the strong measurement at $t=\tau$, the event $x(\tau)=X_{\tau}$ \textit{can} be assigned a well-defined probability measure. In d'Espagnat's terminology, we are dealing with a ``proper mixture''. In short: Owing to the pointer’s entanglement with the particle, determinism \textit{doesn't} imply $\mathbb{E}(y|x(\tau)=\hat{T}_{\tau}\kappa)=\mathbb{E}(y|x(0)=\kappa)$. The initially auspicious argument for (COR) therefore fails.

From its failure, we gain also a wider-reaching insight: unless (at $t=0$) the strong measurement is \textit{actually} performed (unlike in Wiseman’s measurement protocol), the conditional probabilities $\mathbb{P}(y|x(0)=\kappa)$ (or equivalently: their associated expectation values) aren't objectively defined --- \textit{if} one adopts their usual definition in terms of post-selection. Strictly speaking, the \textit{un}realised measurement renders $\mathbb{P}(y|x(0)=\kappa)$, thus defined, meaningless.\footnote{In this proviso, one might descry a possible loophole. Why not define the prerequisite probability measures $\mathbb{P}(y \& x(0)=\kappa)$ and $\mathbb{P}(y|x(0)=\kappa)$ \textit{indirectly} --- via their respective \textit{later} states? That is, instead of the direct definition via post-selection at $t=0$, one might \textit{stipulate} the following probability measures (cf. \citealt{steinberg1995conditional}) 

\begin{equation}
\label{alternative conditional probability 1}
    \mathbb{P}(x(0)=\kappa):=\mathbb{P}(x(\tau)=X_{\tau}), 
\end{equation} and

\begin{equation}
\label{alternative conditional probability 2}
     \mathbb{P}(y\& x(0)=\kappa):=\bra{\Psi( X_{\tau},y,0)\hat{U}_{\tau}^{\dagger}}\ket{\hat{U}_{\tau}\Psi( X_{\tau},y,0)}=\mathbb{P}(y\& x(\tau)=X_{\tau}). 
\end{equation}

The conditional probability densities then are \textit{definitionally} identical with the weak value:

\begin{equation}
\label{defining shift operatior conditional probability}
    \mathbb{P}(y|x(0)=\kappa)\overset{def}{=}  \frac{\mathbb{P}(y \& x(0)=\kappa)}{\mathbb{P}(x(0)=\kappa)} := \frac{\mathbb{P}(y \& x(\tau)=\hat{T}_{\tau}\kappa)}{\mathbb{P}(x(\tau)=\hat{T}_{\tau}\kappa)}=\mathbb{P}(y|x(\tau)=X_{\tau}).
\end{equation}

With this definition, the calculation in \ref{displacement} now uniquely determines the shift operator to be that of standard dBBT. That is: given the indirect definitions \ref{alternative conditional probability 1} and \ref{alternative conditional probability 2}, for this shift operator --- and \textit{only} for it --- (COR) holds.

It may now look, as though Definitions \ref{alternative conditional probability 1} and \ref{alternative conditional probability 2} are indeed the most natural alternatives to the (unavailable) direct definition of the conditional probabilities. Still, nothing invariably \textit{forces} us to resort to any such alternative indirect definition: one may just decline to content oneself with anything but the bona fide direct ones. In light of the comments on entanglement, such an option is in fact quite plausible: one just ought to swallow the fact that the conditional probability fails to be definable.

If one opts for this stance, one will dismiss the operational velocity \ref{operational velocity} as an arbitrary (non-factual) stipulation --- rather than a representation of an objective, physical feature (viz. the particles' real velocity). By the same token, that it coincides with that of standard dBBT, thereby becomes a definitional artefact --- denuded of factual content.

Suppose, however, that one accepts the alternative indirect definitions Equation \ref{alternative conditional probability 1} and \ref{alternative conditional probability 2}. Then, it would still be true that --- by the standard laws of QM --- only one shift operator is consistent with this \textit{definition} --- the shift operator of standard dBBT. That is, in order to employ definitions to describe a deterministic world for which the standard QM formalism is empirically adequate, one must presuppose that standard dBBT (rather than one of its observationally equivalent alternatives) is true. Put differently: given the (operational) Definition \ref{operational velocity} (and the empirical adequacy of QM), (COR) is true, if and only if standard dBBT is true. With regards to the definitions Equation \ref{alternative conditional probability 1} and \ref{alternative conditional probability 2}, we can say: they can only be \textit{adequate}, if standard dBBT is true. That is, they commit us to the presumption of the latter. But that is precisely what renders them problematic stipulations in the present context (cf. \citealt{norton2010hume}).}

\begin{figure}[h]
	\centering
	\includegraphics[width=1\textwidth]{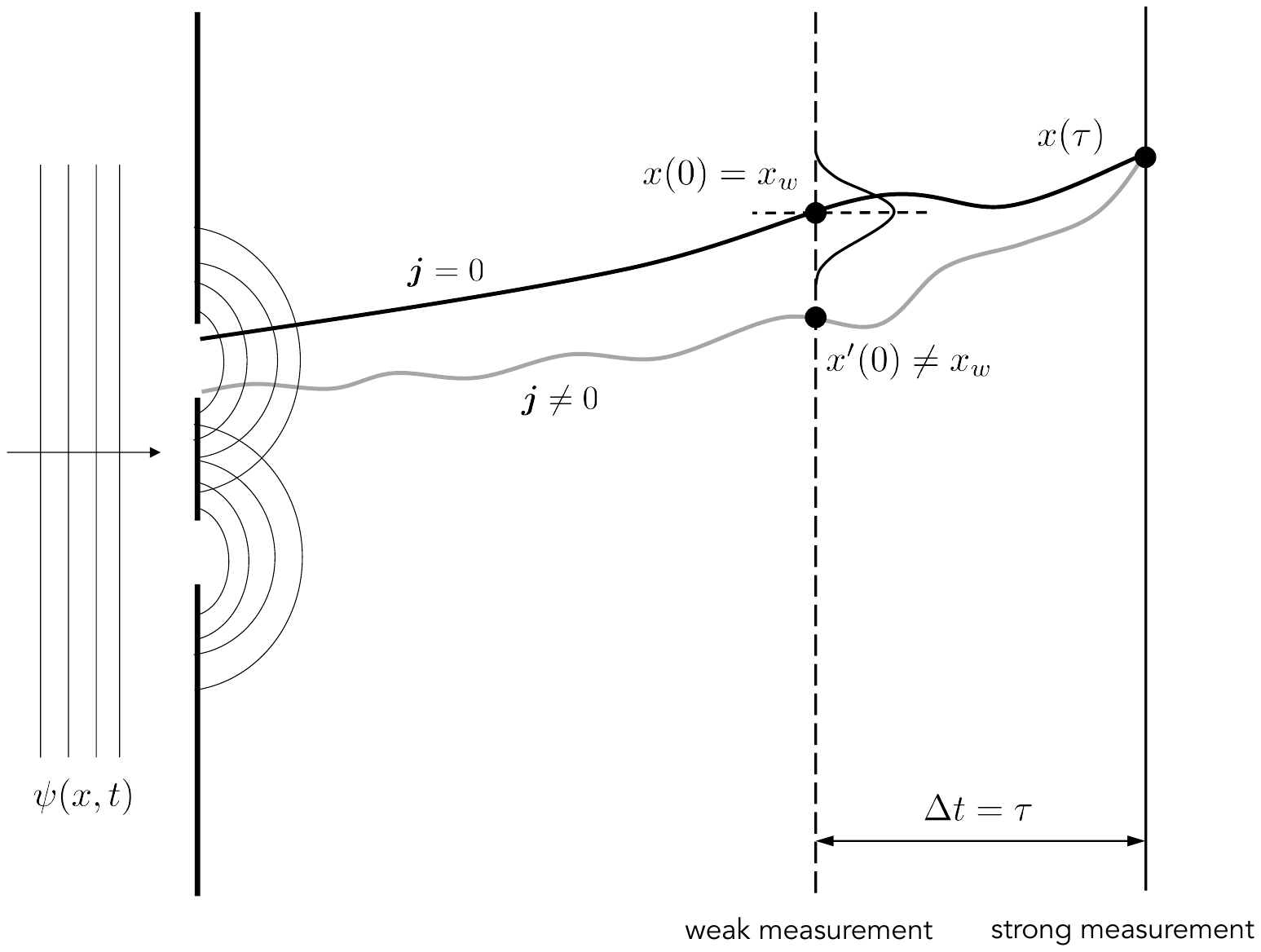}
	\caption{The weak measurement procedure for a given post-selected state $x(\tau)=X_{\tau}$. The weak value is obtained from the distribution on the weak screen. When the velocity field is that of standard dBBT ($\bm{j}=0$), the actual position of the particle $x(0)$ matches the weak value $x_w$. For an alternative guidance equation ($\bm{j}\neq 0$), it doesn't: the particle crosses the weak screen at a point $x'(0)$, other than the weak value. This shows that depending on which guidance equation one chooses, the weak value needn't yield the actual position of the particle at time $0$.}
	\label{fig:correspondence}
\end{figure}

No \textit{independent} reasons have been given so far for believing that (COR) is true, though.
(Conversely, the lack of independent reasons for standard dBBT (rather than any of its non-standard variants), especially in light of its empirical underdetermination, was our major motivation for applying weak velocities in the context of de Broglie-Bohmian theories.)  Consequently, counterexamples to (COR) abound --- and are perfectly familiar: \textit{any} non-standard variant of dBBT of the type of Equation \ref{altguidanceequation} (i.e. with non-vanishing, divergence-free vector field $\mathbf{j}$). In them, the particle's trajectory generically crosses the weak screen at a point \textit{distinct} from what the weak velocity measurements would make us believe. Figure \ref{fig:correspondence} illustrates this.

Nothing compels us --- even if sympathetic to the overall de Broglie-Bohmian framework --- to regard the outcome as \textit{truly} representing the actual position of the particle at time $t$. It's just unknown: it could have traversed \textit{any} location within the support of the Gaussian wave function, centred around the weak value. Still, the operationally defined velocity (obtained from averaging) wouldn't change: as long as the Born Rule and the validity of the Schrödinger Equation --- the only prerequisites for deriving the result Equation \ref{gradient weak velocity} --- hold, its value remains the same. (In this sense, any guidance equation of the type of Equation \ref{altguidanceequation}, is compatible with Wiseman's operationally defined velocity)  

Absent an independent argument for the correspondence between weakly measured and actual positions (i.e. COR), it remains unclear what --- if anything --- Wiseman's operational velocity \ref{operational velocity} signifies ontologically.

By na\"ively generalising $C_0$ to $C_t$, one neglects the relevance of time in the present setup: it matters both when the weak measurement interaction occurs, and \textit{when one post-selects}. If both happen at the same time, the weak position value indeed corresponds to the particle's actual position at time $t=0$. If, however,  some time $\tau$ elapses between interaction and post-selection, generically this is no longer the case. 

It's instructive to rephrase this result: the assumption $C_t$, necessary for the correspondence of weak and actual velocities, is in fact equivalent --- in virtue solely of the quantum mechanical formalism and the supposition of deterministic differentiable particle trajectories --- to standard dBBT. (First, suppose that $C_t$ is true. Then, the weak velocity measurement yields the actual particle velocities. Wiseman's operationally defined velocity \ref{operational velocity} uniquely picks out a guidance equation --- that of standard dBBT. Conversely, suppose standard dBBT to be true. A weak velocity measurement then discloses the actual particle velocities. Thus, $C_t$ holds true.) 

In conclusion: Here, we argued that a particle's weak velocity coincides with its actual velocity (provided one is wiling to attribute deterministic, differentiable paths to the particles), if and only if standard dBBT is true. But this coincidence is a sine qua non for deploying weak velocity measurements in \textit{support} of standard dBBT. To attempt to do so --- absent independent arguments for the reliability of weak velocity measurements --- would one thus incur circularity.  

This analysis permits us to evaluate two verdicts on the significance of weak velocity measurements for standard dBBT, found in the literature. Let's start with Dürr, Goldstein and Zangh\`i's claim that they enable genuine measurements.

\subsection{Weak measurements as genuine measurements?}
\label{DGZgenuine}

The foregoing analysis sheds light on a recent claim by Dürr, Goldstein, and Zangh\`i (\citeyear{Durr2009}). These authors (henceforth abbreviated as ``DGZ'') aver that Wiseman’s measurement protocol for weak velocities allows ``in a reasonable sense, a \textit{genuine} measurement of velocity'' in standard dBBT (ibid., pp.~1025, DGZ’s emphasis). Such a statement, we maintain, is misleading. DGZ themselves identify a condition as crucial for their claim. This identification, too, we deem the source of further potential confusion. The crucial --- but in DGZ's account \textit{tacit} --- condition for weak velocity measurements to be reliable, as we saw in the previous sections, is (COR). But (COR) is equivalent to assuming the standard form of the guidance equation. The essential equivalence between (COR), and  dBBT's standard guidance equation impinges upon the significance of weak measurements for dBBT: whether we regard weak velocity measurements as enabling genuine measurements of the particle's actual velocity is essentially equivalent to an \textit{antecedent} commitment to standard dBBT. Pace DGZ, this curtails the significance of weak velocities as genuine. Yet, albeit misplaced in the context of weak measurements, DGZ's (misleadingly) identified crucial condition might open up a potentially illuminating perspective on standard dBBT.

DGZ assert that weak velocity measurements, as realised by Wiseman’s measurement protocol, constitute real measurements in standard dBBT (cf. \citealt[Section 3.7]{Durr2004}). What is more, in his authoritative review of dBBT \cite[Section~4]{goldstein-sep-qm-bohm} writes: ``In fact, quite recently \cite{kocsis2011observing} have used weak measurements to reconstruct the trajectories for single photons `as they undergo two-slit interference, finding those predicted in the Bohm-de Broglie interpretation of quantum mechanics' '' (cf. \citealt[pp.~142]{durr2018verstandliche} for a similar statement). DGZ are aware of the fact that such a claim needs an additional assumption; they (as we'll show) misidentify that ``crucial condition'' \citep[p.~1026, 1030]{Durr2009}.  

Before adverting to DGZ's declaration of weak velocity measurements as genuine, a repudiation is apposite of the claim that such weak velocity measurements have actually been performed, \textit{in accordance with dBBT's predictions}.

Figure \ref{fig:Kocsis} displays the weak velocities measurements ascertained in Kocsis et al.'s double-slit experiment. Indeed, they qualitatively tally with the trajectories of standard dBBT (cf., for instance, Figure 5.7 in \citealt[p.~184]{holland1995quantum}). Still, \textit{nothing} immediately follows from that regarding the status of standard dBBT (see also \citealt{flack2014weak,flack2016weak}; \citealt[p.~181]{bricmont2016making})!  Kocsis et al.'s experiment has been performed for (massless) \textit{photons}. Standard dBBT, however, is a non-relativistic quantum theory for massive particles: as such, it can't handle photons.\footnote{The treatment of photons within field-theoretic \textit{extensions} of dBBT, capable of dealing with photons (or bosons, more generally), is a delicate matter, outside the present paper's ambit. We refer the interested reader to e.g. \citealt[Chapter~11]{holland1995quantum} and \citealt[Chapter~10]{durr2012quantum} (also for further references).} Kocsis et al.'s experiment hence has no direct bearing on dBBT's status.\footnote{Rather than the trajectories of \textit{individual} photons, \citealt{flack2014weak} and \citealt{flack2016weak} have argued that Kocsis et al.'s experiments measure mean momentum flow lines. 

This interpretation (as we saw in Equation \ref{gradient weak velocity}) has a counterpart in weak velocity measurements of the electrons of the present setup: per se, the weak velocity measurements only allow experimental access to the gradient of the wave function's phase. (This view on weak velocity values remains neutral, though, vis-à-vis any interpretation of the wave function. In particular, it's not necessarily committed to a statistical/ensemble interpretation.)}     

Now to DGZ's main claim, as we understand it: that for a coherent application of weak velocity measurements to the Bohmian framework as reliable velocity measurements, one needs an assumption on the disturbance of actual velocities is needed. Only standard dBBT, so the story goes, has this feature. In turn it appears that weak velocity measurements can constitute genuine measurements of the particle's actual velocities only in standard dBBT.

DGZ’s considerations seem to \textit{start from} the reliability of weak velocity measurements; they are predicated on (COR). DGZ (correctly) state that only standard dBBT is consistent with that. As the ``crucial condition'', responsible for that result, they identify a characteristic feature of standard dBBT's velocity field.
\newpage

\begin{itemize}
    \item[\textbf{(SPE)}]  
    \textit{Whenever the particle-\textit{cum}-pointer compound system has the form $\psi(x)\otimes\phi(y-x)$, the particle’s velocity field $v$ (conceived of as a function of the compound system’s wave function $\psi\otimes\phi$) is supposed to depend only on the particle’s wave function $\phi$:}
    $v[\psi\otimes\phi]=v[\phi]$.
\end{itemize}

We‘ll dub this condition ``separability of particle evolution'' (SPE). It uniquely singles out standard dBBT \citep[Section~4]{Durr2009}.

DGZ’s mathematical proof of this latter claim is beyond dispute. Their identification of (SPE) as a \textit{physically} essential condition, however, is wrong-headed: (SPE) in fact plays no obvious role in the attempt to exploit weak velocity measurements for standard dBBT (see §\ref{dBBTweak} and §\ref{secton when do weak and actual velocities coincide}): nowhere is it invoked explicitly. Moreover, it remains elusive how (SPE) \textit{could} enter that analysis: (SPE) is an \textit{exact} equality, postulated to hold, whenever the composite particle-pointer wave function is factorisable. By contrast, DGZ‘s decisive equations (viz. (21) and (22) in their paper) are only approximations, valid at $t=\tau$. Their terms linear in $\tau$ \textit{don't} take a factorisable form (nor do they vanish). Not even at $t=0$ is the pointer-particle wave function factorisable. Hence, (SPE) doesn't seem to be applicable from the outset. To call (SPE) ``crucial'' --- understood as \textit{directly} responsible --- for the reliability of weak velocity measurements in dBBT muddies the waters: \textit{it‘s solely in virtue of (SPE)‘s essential equivalence with standard dBBT} that (SPE) is relevant at all. That (SPE) singles out standard dBBT is salient of the (mathematical) form of the standard guidance equation: the latter is uniquely characterised by the factorisation of velocities at $t=0$, as asserted by (SPE). 

As a result, only because (COR) presupposes standard dBBT, \textit{and} because the latter is essentially equivalent to (SPE) (recall our remark at the end of §\ref{secton when do weak and actual velocities coincide}), is (SPE) ``crucial'' --- in the sense of necessarily satisfied for (COR) to hold. In short: (COR), (SPE) and standard dBBT’s guidance equation are essentially equivalent. That is:
\begin{align}
    (COR) \wedge (DIF) \wedge (DET) \nonumber& \Longleftrightarrow \text{dBBT's standard guidance equation} \\\nonumber
    & \Longleftrightarrow (SPE) \wedge (DIF) \wedge (DET), \nonumber
\end{align} where (DET) and (DIF) denote the assumption of deterministic and differentiable particle trajectories, respectively. 

For weak velocity measurements to reveal the particles' actual trajectories (assuming determinism and differentiability, that is) --- i.e. for weak velocity measurements to be reliable --- (COR) \textit{not} (SPE) --- is the crucial condition that must be satisfied: without it, the counterfactual $C_t$ no longer holds (recall \ref{secton when do weak and actual velocities coincide}); the particle's later positions can't be inferred from the weak measurements. In particular given (COR)'s essential equivalence with standard dBBT or (SPE), this means that if weak velocity measurements are reliable, (SPE) needn't be assumed separately: it's implied by (COR). We thus reject DGZ's identification of (SPE) as the crucial condition for the reliability of weak measurements. Pace DGZ, one might hence baulk at calling them genuine in a sufficiently \textit{robust} sense. Unless \textit{independent} reasons for (SPE), (COR) or standard dBBT are forthcoming, weak velocity measurements lack epistemic significance for gauging the status of dBBT. The analysis of weak measurements in a de Broglie-Bohmian framework \textit{doesn't} rely on (SPE). DGZ are right, however, in observing that if standard dBBT is true, weak measurements are reliable (i.e. weak position values and actual position values coincide).  

DGZ’s purely mathematical result --- the equivalence of (SPE) and standard dBBT --- hints at an alluring possibility (completely independently of weak measurements): it \textit{might} serve as a prima facie interesting avenue for justifying (or, at least, motivating) standard dBBT. Underlying (SPE) seems to be the hunch that for particle-pointer systems with separable (factorisable) quantum states, the particle is supposed to be guided exclusively by the particle’s wave function --- not by that of the pointer. More generally, due to (SPE), whenever a quantum system is prepared as separable, the dynamics for the particles of one subsystem doesn't depend on the quantum state of other subsystem(s).\footnote{This is somewhat reminiscent of so-called Preparation Independence, a key assumption in the Pusey-Barrett-Rudolph Theorem (see e.g. \citealt[sect. 7]{Leifer2014extended}; specifically for the theorem in the context of standard dBBT, see \citealt{drezet2014pbr}). Roughly speaking, Preparation Independence asserts that in hidden-variable theories, the so-called ``ontic states'' (i.e. the states represented by the two systems' ``hidden'' variables), should be statistically independent, if their joint quantum state is separable. For hidden-variable theories, this looks like a natural desideratum: it expresses how the separable systems' independence at the quantum level (cf. e.g. \citealt{howard1989holism,howard1992locality}) percolates to (i.e. constrains) the level of the more fundamental level of the hidden-variables (cf. \citealt[sect. 7.3]{Leifer2014extended}).

There exists a critical difference between Preparation Independence and (SPE): the former makes claims about the hidden-variable \textit{states} (in the present case: the particles' positions) and their statistically independent distribution; SPE, by contradistinction, makes a claim about their \textit{dynamics} --- its independence of other subsystems' quantum state (see main text). Consequently, one ought to expect justifications for either to differ.

In fact, Preparation Independence doesn't imply (SPE): all variants of dBBT respect the former (due to the Born Rule giving the distribution of the particles' actual positions, see e.g. \citealt{Gao2019pbr}) --- but only standard dBBT satisfies (SPE).}

As a desideratum, (SPE) implements the expectation that the statistical independence at the quantum level percolates to the level of the behaviour (i.e. dynamics) of the hidden-variables: whenever the quantum states of a composite system $A\&B$ are independent, the dynamics of the particles constituting $A$ shouldn't be affected by $B$’s quantum state. One may deem this a plausible (albeit defeasible) heuristic principle for the construction of hidden-variable theories: it aligns the statistical independence of the known (empirically accessible) realm of the quantum formalism (for separable quantum states), and the independence of the unknown  (empirically inaccessible) realm of the (putatively) more fundamental hidden-variables' dynamics. A dynamics respecting this alignment, one might feel, ``naturally'' explains the statistical independence at the coarse-grained quantum level. 

On the other hand, one may well query the status of (SPE). The separability of quantum states is arguably related to their individuation (see e.g. \citealt{howard1985einstein,howard1989holism}; \citealt[Appendix~B3]{brown2005physical}): for composite systems with separable quantum states, subsystems have distinct quantum states. But why deem the individuation of quantum states --- usually construed in this context as encoding statistical, \textit{coarse-grained} properties to which our empirical knowledge seems limited --- relevant for a constraint on the (putatively more) fundamental particle dynamics? Even if the particle and the pointer possess distinct (individual) quantum states, why should it follow that the particle’s dynamics should depend only on the particle’s wave function?

What might seem to suggest that is that SPE encodes a form of locality. (\textit{Standard} (Bell-)locality forbids an action-at-a-distance. The kind of locality enshrined in (SPE) forbids that a particle's dynamics depends on the pointer's quantum state, even if the joint quantum state of the particle and the pointer is separable.) But standard non-locality is a manifest, distinctive feature of dBBT. The type of locality that (SPE) asserts doesn't restore standard locality. What then is it supposed to achieve? We leave the prospects of (SPE) as a potentially promising motivation for standard dBBT to future inquiry.



This section afforded two main lessons. Standard dBBT is mathematically uniquely characterised by a factorisation condition on the velocity field. We argued that DGZ's identification of that condition as ``crucial'' for the reliability of weak measurements was misleading. 2. Weak velocities coincide with the particle’s actual velocities, if and only if standard dBBT is true. It thus remains questionable what argument (if any) weak velocity measurements provide in support of standard Bohmian trajectories or any other Bohmian theory. 

On their own, weak velocity measurements thus don't provide any empirical support for standard dBBT. What about \textit{non}-empirical inferential support, though?

\subsection{Non-empirical support for dBBT?}
\label{grounding}

The main result of Wiseman's original paper can be read as a conditional claim: \textit{if} one adopts his operationally defined velocity, \textit{and} assumes deterministic, differential particle trajectories, the latter is uniquely determined as that of standard dBBT; on this reading, Wiseman remains neutral vis-à-vis this claim's premises --- whether they are plausibly satisfied (or not). Stated thus, Wiseman's stance is impeccable. More exciting, however, would be the prospect of learning something novel about the status of standard dBBT from weak measurements (granting certain background assumptions). We'll now examine such a stronger interpretation of Wiseman's result: as a non-empirical justification of standard dBBT.    

We flesh out three possible variants of such an argument.\footnote{This may be thought of as an eliminative induction (see e.g. \citealt{norton1995eliminative}), where one eliminates from a universe of candidate theories all but one with the help of background assumptions and principles.}

The starting point of the envisioned reasoning will be two tenets, explicitly endorsed by Wiseman:
\begin{itemize}
    \item[(1)] One should construe the weak value in Wiseman’s weak measurement protocol of §\ref{operational velocity} as the average velocity of a large ensemble of particles \citep[sect. 3]{wiseman2007grounding}.
    \item[(2)] Albeit not per se referring to individual particles, this \textit{statistical} property provides a ``justification for [standard dBBT's] law of motion [i.e. the standard guidance equation]'' (ibid., p.~2).
\end{itemize}

According to tenet (1), the weak value, obtained in Wiseman’s setup, by itself corresponds to a real property only of an ensemble of particles --- rather than one na\"ively ascribable to the individual particles:  ``Thus strictly the weak value [...] should be interpreted [...] only as the mean velocity in configuration space --- this noise could be masking variations in the velocity between individual systems that have the same Bohmian configuration $x$ at time $t$.'' (\citealt[p.~5]{wiseman2007grounding}). 

One of the premises in the conditional claim is determinism. With that assumption in place, weak values within a de Broglie-Bohmian framework are plausibly interpreted as first and foremost statistical properties of ensembles, as asserted in (1): formally, weak values are (normalised) transition amplitudes (cf. \citealt{kastner2017demystifying}; pace, \citealt{vaidman1996weak}). Hence, the usual interpretation of probability amplitudes within dBBT as statistical (ensemble) properties applies (see e.g. \citealt[Chapter~3.8]{holland1995quantum}; \citealt[Chapter 9.3]{bohm2006undivided}).\footnote{Outside of the framework of dBBT --- in particular outside of epistemic interpretations of the Born Rule-based quantum formalism --- the interpretation of weak values as ensemble properties is more tenuous (see \citealt{matzkin2019weak}). Here, we'll set aside such quibbles.} 

Tenet (2) purports that \textit{in virtue of this statistical (ensemble) property} dBBT’s standard form ``is preferred over all other on physical grounds'' (\citealt[p.~12]{wiseman2007grounding}). That is, although other velocity fields generate the same (statistically-empirically accessible) mean velocity, we ought to believe that the standard velocity field is true --- rather than any of its alternatives: for Wiseman, (2) serves as a non-empirical rule of inference\footnote{The intended form of non-empirical support is distinct from Dawid's (\citeyear{dawid2013string,dawid2019}) ideas of non-empirical confirmation. Regardless of how convincing one finds Dawid's proposal, it doesn't apply to the present case.}, ``justifying [dBBT's] foundations'' (ibid., p.~12). 

As Wiseman reiterates, no experiment can discriminate between dBBT’s standard velocity field and alternative choices. How then is the envisaged non-empirical justification supposed to work? What undergirds (2)? Three strategies (intimated to some extent by Wiseman and his commentators) spring to mind: \textit{(A)} some variant of operationalism, \textit{(B)} simplicity and/or parsimony, and \textit{(C)} some variant of inference to the best explanation. 

\textit{(A)} The first invokes some form of operationalism in the spirit of \citealt{bridgman1927logic}. In its crudest form, it demands that all theoretical quantities be operationalisable: there must exist suitable measurement instructions for them. Yet, operationalism ``[...] is nowadays commonly regarded as an extreme and outmoded position‘‘ (\citealt{chang-sep-operationalism}, also for a compilation of the arguments against operationalism). We’ll therefore not discuss it further. 

Perhaps an attenuated form fares better --- one according to which (ceteris paribus) it’s merely \textit{desirable} that theoretical quantities be operationalisable. Wiseman seems to cherish the desideratum that ``the [Bohmian particle] dynamics are deterministic, and that the velocity-field of the [hidden variable, i.e. the particle positions] should be na\"ively observable [...]''. But what would buttress such a desideratum? In particular, why believe that a theory that satisfies it is more likely to be true than empirically equivalent rival theories that don't? 

\textit{(B)} A second strategy (expressly disavowed by Wiseman) might turn on simplicity. Wiseman's operational definition, on this line of thought, should be regarded as distinguished --- as particularly simple. Even if we set aside both Wiseman’s concern that ``simplicity is not a property that can be rigorously defined'' \citep[p.~9]{wiseman2007grounding}, and the problematic assumption that simplicity is truth-conducive, an appeal to simplicity isn't promising: simplicity and postulating that individual particle trajectories coincide with their statistical averages are unrelated. Although intuitively it may prima facie appear \textit{simple}, if the individual trajectories are chosen so as to coincide with their statistical averages, the precise sense of simplicity turns out to be elusive: neither the theory’s qualitative nor its qualitative parsimony are affected by that choice. That is, neither new or additional kinds/types of entities are introduced or eliminated in the theory’s ontology, nor is the overall number of entities multiplied or reduced.

To appeal to parsimony would likewise be of no avail: neither in terms of quantitative (i.e. with respect to numbers of individual entities postulated) nor qualitative (i.e. with respect to numbers of types or kinds postulated) parsimony does such a postulate seem privileged. 

\textit{(C)} A third attempt to defend (2) might appeal to an Inference to the Best Explanation (IBE) (see e.g. \citealt{lipton2003inference}; \citealt[Chapter~4]{bartelborth2012erkenntnistheoretischen}): standard dBBT, on this view, provides the best explanation for the observational facts in Wiseman’s protocol. 

Again, let’s grant that IBEs are generically justifiable (pace e.g. \citealt[Chapter~2]{van1980scientific}; \citealt[Part~ II]{van1989laws}). Yet, in light of the foregoing comments on parsimony and simplicity, it’s opaque in which sense standard dBBT could explain (or help us understand) the empirical phenomena in any \textit{better} way than versions with non-standard velocity fields; both are \textit{equally} capable of accommodating the empirical phenomena.  

A variant of this appeal to an IBE\footnote{The variant of an IBE is known as ``Inference to the \textit{Loveliest} Explanation'' (see  \cite{lipton2003inference}, passim). Instances of the latter are supposed to provide the best \textit{understanding} of the phenomena. Even if one grants the controversial assumption that loveliness confers also likeliness to be true, the objection in the main text carries through: in no palpable, non-arbitrary way does it strike us as particularly ``lovely'', if the particles' mean velocity coincides with that of their individual ones. It certainly doesn't \textit{enhance} our understanding.

Dürr et al. might retort that the conditions of their uniqueness proof for standard dBBT's guidance equation, mentioned in §\ref{section underdetermination}, bring standard dBBT's loveliness to the fore. In response there is little more we can do than profess our differing aesthetic predilections.}, found in the literature, fixates on Wiseman’s emphasis of the allegedly natural character of his proposal to operationally define velocities via weak values: ``(Standard dBBT) delivers thus the most natural explanation of the experiments described''(\citealt[p.~145]{durr2018verstandliche}, our translation). 

Three reasons militate also against this view. First, the intended notion of a natural explanation is to our minds vague. Hence, it’s difficult to fathom its argumentative force. At best, it seems an aesthetic criterion. As such, its suitability for assessing theories is suspect (cf. \citealt{ivanova_forthcoming,ivanova2017aesthetic, hossenfelder2018lost}).

Secondly, in light of the highly \textit{unnatural} consequences of the same reasoning in other contexts, one may well debate whether Wiseman's operationally defined velocity is indeed natural after all. \citet[p.~223]{aharanov2005quantum} --- presumably against the authors' intentions --- summarise the generic ``unnaturalness" of weak values: ``weak values offer intuition about a quantum world that is freer than we imagined --- a world in which particles travel faster than light, carry unbounded spin, and have negative kinetic energy.''

Thirdly, and quite generally, in §\ref{secton when do weak and actual velocities coincide} and §\ref{DGZgenuine} we have seen that in the present case the allegedly natural explanation would at any rate be deceitful: one mustn't \textit{na\"ively} take it for granted that they reveal the actual particle positions. \citet{leavens2005weak} draws attention to the fact that under certain experimental circumstances ``[...] there is no possibility of the weak value [...] reliably corresponding in general, even on average, to a successfully post-selected particle being found near (the weak value) at time $t=0$ when the impulsive weak position measurement begins and being found near (the post-selected value) an instant after it ends'' (p. 477). 

The perils of na\"ive (i.e. literal) realism about weak position values are drastically demonstrated in the so-called Three-Box-Paradox (\citealt{aharonov1991complete}; \citealt[Chapter~16.5]{aharanov2005quantum}; \citealt{maroney2017measurements}). Imagine a particle and three boxes, labelled $A$,$B$, and $C$. Let the particle‘s initial state be 

\begin{equation}
    \ket{\psi_i}= \frac{1}{\sqrt{3}}(\ket{A}+\ket{B}+\ket{C}),
\end{equation}

where $\ket{A}$ denotes the state in which the particle is in box $A$, and similarly, $\ket{B}$ and $\ket{B}$. For its final state, on which we’ll post-select, choose

\begin{equation}
    \ket{\psi_f}= \frac{1}{\sqrt{3}}(\ket{A}+\ket{B}-\ket{C}).
\end{equation}

Via the definition of weak values (see \ref{weak values appendix}), one then obtains the resulting weak values for the projectors onto state $i\in {A,B,C}$, $\hat{P}_i:=\ket{i}\bra{i}$:

\begin{align}
\langle\hat{P}_A\rangle_w&= 1\nonumber\\
\langle\hat{P}_B\rangle_w&= 1\nonumber\\
\langle\hat{P}_C\rangle_w&= -1.
\end{align} 

If one were to believe that weak values invariably reveal the real positions of particles, one would have to conclude that box $C$ contains $-1$ particle! Within the ontology of dBBT (in any of its variants), this is an absurd conclusion: particles in dBBT either occupy a position or they don’t; the respective position projectors take values only in $\{0,1\}$.

Consequently, it's imperative that adherents of dBBT be wary of interpreting weak values as real position values without qualification. Our analyses in §\ref{secton when do weak and actual velocities coincide} and §\ref{DGZgenuine} underscore this: the reliability of weak position (or velocity) measurements is a non-trivial (and generically \textit{false}) assumption.

In conclusion, our hopes were dashed that the velocity measurement in Wiseman's protocol supports dBBT in any robust, non-empirical sense. Neither the alleged merits of operationalisability per se nor considerations of simplicity or parsimony warrant it. An IBE proved implausible. Unqualified realism about weak position values inevitably conflicts with dBBT's default ontology. 

We are thus left with at best a considerably weaker position, one close to Bricmont's (\citeyear[p.~136]{bricmont2016making}): ``[Weak velocity measurements via Wiseman’s protocol] (are) not meant to `prove' that the de-Broglie-Bohm theory is correct', because other theories will make the same predictions, but the result is nevertheless suggestive, because the predictions made here by the de Broglie-Bohm theory is [sic] very natural within that theory [...].''

Understanding that suggestiveness and ``naturalness" possess scant epistemic or even non-subjective import, we concur. With such a verdict, however, one has  relinquished the initial hope that weak measurements per se have a fundamental bearing on whether standard dBBT or one of its alternative versions are true. 

\section{Conclusion}
\label{conclusion}

Let's recapitulate the findings of this paper. 

We started from the empirical underdetermination of dBBT’s guidance equation. It poses a impediment to insouciant realism about the particles' trajectories, postulated by standard dBBT. We scrutinised whether Wiseman’s measurement protocol for weak velocities is able to remedy this underdetermination by empirical or non-empirical means. Our result was negative. We elaborated that the reliability of weak velocities --- the fact that they coincide with the particles' real velocities --- presupposes standard dBBT. For non-standard versions of dBBT, its presumption is generically false. Hence, weak velocity measurements don’t qualify as evidence or confirmation in favour of the velocity field, postulated by standard dBBT. Weak velocity measurements thus don't allow for genuine measurements in any robust sense (at least given the present knowledge). Finally, we critiqued an interpretation of Wiseman's measurement protocol as a non-empirical argument for standard dBBT in terms of alleged theoretical virtues. Even if one grants the questionable appeal to some popular virtues, it remains equivocal that in the context of weak velocity measurements standard dBBT actually exemplifies them. Most importantly, the 3-Box Paradox demonstrated the dangers of \textit{any} na\"ive realism about weak \textit{position} values. 

In conclusion, our paper has, we hope, elucidated the status of weak velocity measurements in two regards. On the one hand, they are indubitably an interesting application of QM in a novel experimental regime (viz. that of weak pointer-system couplings). They allow us to empirically probe the gradient of the system's wave function --- irrespective of any particular interpretation of the quantum formalism. 
On the other hand, however, with respect to the significance of weak velocity measurements, we proffered a deflationary account: per se, weak velocity measurements shed no light on the status of standard dBBT. In particular, on their own, they don’t provide any convincing support --- empirical or non-empirical --- for standard dBBT over any of its alternative versions.

\section{Appendix: Weak measurements and weak values}

\label{Appendix}
Methods of weak measurement have opened up a flourishing new field of theoretical and experimental developments  (see e.g. \citealt{aharanov2005quantum}; \citealt{TamirCohen2013}; \citealt{svensson2013pedagogical}; \citealt{dressel2014colloquium}. Broadly speaking, weak measurements generalise strong measurements in that the final states of measured systems need no longer be eigenstates. In this appendix, we'll first provide a concise overview of weak measurements (§\ref{strong measurements appendix}). In particular, we'll expound how they differ from the more familiar strong ones . In §\ref{weak values appendix}, we'll introduce notion of a weak value. 

\subsection{Strong versus weak}
\label{strong measurements appendix}

Strong or ideal measurements are closely related to the conventional interpretation of the Born Rule. Consider a quantum system $\mathcal{S}$ and a measuring device $\mathcal{M}$ with Hilbert spaces $\mathcal{H_S}$ and $\mathcal{H_M}$, respectively. The Hilbert space of the total system is $\mathcal{H}= \mathcal{H_S}\otimes\mathcal{H_M}$. The system be in a normalized state $\ket{\psi}$ before the measurement. We are interested in measuring an observable $A$ represented by the self-adjoint operator $\hat{A}$, which has a complete and orthonormal eigenbasis $\{\ket{c_i}\}$. In that basis the system's state reads $\psi=\sum\limits_i\alpha_i\ket{c_i}$ for some $\alpha_i$.  Furthermore, we assume for simplicity the eigenstates are non-degenerate, i.e. have distinct eigenvalues. The only possible outcome of a strong measurement on this system is one of the eigenstates $\ket{c_i}$. The corresponding probabilities to observe $\ket{c_i}$ are
\begin{equation}
p_i=|\bra{c_i}\ket{\psi}|^2=|\alpha_i|^2.
\end{equation} 
After the measurement was performed the system ends up in the final state $\ket{c_i}$. This procedure is known as the \textit{von Neumann measurement} (cf., for example, see the reprint \citep{von2018mathematical}).  

In a weak measurement the interaction of system and measurement device is modelled quantum mechanically with the pointer device as an ancillary system on which a strong measurement is performed after the interaction. That is, assume that system and pointer interact via a von Neumann Hamiltonian 
\begin{equation}
\hat{H}=g(t)\hat{A}\otimes \hat{P}_M,
\end{equation}, where $\hat{P}_M$ is conjugate to the pointer variable $\hat{X}_M$, i.e. $[\hat{X}_M, \hat{P}_M]=i\hbar$. As before $\hat{A}$ is the quantum operator of the observable to be measured, and $g(t)$ a coupling constant satisfying $\int\limits_0^T g(t) dt =1$. For simplicity, take a single qubit prepared in initial state
\begin{equation}
\ket{\psi}= \sum\limits_i \alpha_i\ket{c_i}= \alpha\ket{0}+\beta\ket{1}.
\end{equation} We stipulate the eigenvalues of $\hat{A}$ are $\hat{A}\ket{0}=\ket{0}$ and $\hat{A}\ket{1}=-\ket{1}$. Suppose that the pointer that is to be coupled weakly to the qubit will initially be in a Gaussian ready state with spread $\sigma$ peaked around $0$, i.e.
\begin{equation}
\ket{\varphi}=\int\varphi(x)\ket{x}dx=\int N e^{-\left(\frac{x}{2\sigma}\right)^2}\ket{x}dx,
\end{equation} with $N$ a normalization factor. During the unitary interaction, the total initial state $\ket{\Phi(0)}=\ket{\psi}\otimes\ket{\varphi}$ of system and pointer evolves according to Schrödinger's equation: 
\begin{align}
\label{qubit evolution}
\ket{\Phi(T)}&= e^{-\frac{i}{\hbar}\int\limits_0^T \hat{H} dt}\ket{\Phi(0)}\\ \nonumber
&=  e^{-\frac{i}{\hbar}\hat{A}\otimes \hat{P}_M}\ket{\Phi(0)}\\\nonumber
&= e^{-\frac{i}{\hbar}\hat{A}\otimes \hat{P}_M}\left(\alpha\ket{0}+\beta\ket{1}\right)\otimes \int\varphi(x)\ket{x}dx\\ \nonumber
&= N\int \left(\alpha e^{-\left(\frac{x-1}{2\sigma}\right)^2}\ket{0}+\beta e^{-\left(\frac{x+1}{2\sigma}\right)^2}\ket{1}\right)\otimes\ket{x}dx.
\end{align} Recall that the momentum operator acts as a shift operator ($e^{-\frac{i}{\hbar}aP_M}\varphi(x)=\varphi(x-a)$). If the Gaussian peaks are narrowly localized and non-overlapping (to a good approximation), one can infer the state of the system from the pointer measurement. However, for weak measurements the Gaussians are assumed to widely spread over the pointer variable. The measurement outcome of the pointer is therefore consistent with the system being in states that are not eigenstates of the operator. This is read off from Equation \ref{qubit evolution}. If, say, the pointer ends up at position $0$, for example, we recover the initial state $\ket{\psi}$ up to an overall factor. The two Gaussian amplitudes reduce to the same value. 

For arbitrary systems with finite Hilbert space, the interaction generalises to 
\begin{equation}
\label{endstate}
\ket{\Phi(T)} = \sum\limits_i \alpha_i\ket{c_i}\otimes\int\varphi(x-a_i)\ket{x}dx,
\end{equation} where $a_i$ are the eigenvalues of the measurement operator $\hat{A}$. For simplicity, the free evolution Hamiltonian of system and pointer has been omitted; it would only give rise to additional total phases.  

So far the measurement scheme was standard. In Equation \ref{endstate} no weakness is involved in particular. It becomes a weak one if the initial state of the pointer variable $X_M$ has a large spread $\sigma$. That is, the result of (strong) measurement on the pointer is not a projection onto eigenstates of the system. 

\subsection{Post-selection and two-vector-formalism}
\label{weak values appendix}

We may now introduce the notion of a weak value. A weak value of an observable $\hat{A}$ is the result of an effective interaction with the system in the limit of weak coupling and a subsequent post-selection. Coming back to the simple case of the qubit, if the state in Equation $\ref{qubit evolution}$ is post selected on $\ket{0}$, for instance, the pointer ends up in a Gaussian lump centered around $1$. Similarly, conditioned on $\ket{1}$ the pointer is centered around $-1$, as one would expect from a strong measurement as well. Depending on the choice of the post-selected state, however, the pointer states are ``reshuffled'' and can be concentrated around mean values that can be far away from the eigenvalues of the observable $\hat{A}$. In the limit of large standard deviation $\sigma$ the distribution is again Gaussian though. For post-selecting $\ket{+}:=\frac{1}{\sqrt{2}}(\ket{0}+\ket{1})$, for example, the distribution of the measurement device peaks around 
\begin{equation}
a_w= \frac{\alpha-\beta}{\alpha+\beta}.
\end{equation} This is easily obtained by observing
\begin{equation}
\ket{+}\bra{+}\otimes  \mathds{1}  \ket{\Phi(T)}= \ket{+}\otimes\frac{N}{\sqrt{2}}\int\left(\alpha e^{-\left(\frac{x-1}{2\sigma}\right)^2}+\beta e^{-\left(\frac{x+1}{2\sigma}\right)^2}\right)\ket{x}dx.
\end{equation} In the weak limit $\sigma\gg 1$ this gives
\begin{equation}
\approx \ket{+}\otimes\frac{N}{\sqrt{2}}\int(\alpha+\beta) e^{-\left(\frac{x-\frac{\alpha-\beta}{\alpha+\beta}}{2\sigma}\right)^2}\ket{x}dx.
\end{equation}

Importantly, the measurements on the pointer and the ones to find a post selected state are \textit{strong measurements} in the sense defined above. For arbitrary post-selection on a final state $\ket{\psi_f}$ the state of the total system evolves according to

\begin{equation}
\ket{\psi_f}\bra{\psi_f}e^{-\frac{i}{\hbar}\int\limits_0^T \hat{H} dt}\ket{\psi_i}\otimes\ket{\varphi}.
\end{equation} Since the spread $\sigma$ is large, the interaction Hamiltonian, which produces a shift in the pointer's wave function, can be effectively approximated by $e^{-\frac{i}{\hbar}\int\limits_0^T \hat{H} dt} \approx 1- \frac{i}{\hbar}\hat{A}\otimes\nobreak\hat{P}_M T$. Thus, the final state reads

\begin{align}
&\approx \ket{\psi_f}\otimes\bra{\psi_f}\ket{\psi_i}\left(1- \frac{i}{\hbar}a_w\hat{P}_M T\right)\ket{\varphi} \nonumber\\
&\approx \ket{\psi_f}\otimes\bra{\psi_f}\ket{\psi_i}e^{-\frac{i}{\hbar}a_w\hat{P}_M}\ket{\varphi},
\end{align} where

\begin{equation}
\label{weak value}
\langle\hat{A}_w\rangle:= a_w= \frac{\bra{\psi_f}\hat{A}\ket{\psi_i}}{\bra{\psi_f}\ket{\psi_i}}
\end{equation} the salient quantity of the weak value of the observable operator $\hat{A}$. That is, after many runs, the pointer's average position is $a_w$\footnote{There are cases in which $a_w$ is complex. Then, besides the position, the momentum is shifted too}. In other words, $\ket{\varphi}$ experiences the shift $\varphi(x) \mapsto \varphi(x-a_w)$. Note that the probability amplitude to obtain $\ket{\psi_f}$ in the post-selection is $p=|\bra{\psi_f}\ket{\psi_i}|^2$. If the initial and final state of $S$ are nearly orthogonal, the measurement may require many runs to find $a_w$ as the post selected state occurs only rarely. If there is time evolution of the target system between the weak interaction and the final measurement of $\bra{\psi_f}$, then the expression would include $\bra{\psi_f}U$, where $U$ the unitary evolution operator:
\begin{equation}
\langle\hat{A}_w\rangle:= a_w= \frac{\bra{\psi_f}U\hat{A}\ket{\psi_i}}{\bra{\psi_f}U\ket{\psi_i}}.
\end{equation} For a derivation, we refer the interested reader to literature.

\subsection{Weak velocity and the gradient of the phase}
\label{weak velocity and the gradient of the phase}
We can manipulate the definition of the operationally defined weak velocity to give us the velocity of the guidance equation of standard dBBT.  That is, for the unitary evolution $\hat{U}(\tau)=e^{-i\hat{H}\tau/\hbar}$ during time $\tau$ (with the non-relativistic Hamiltonian of a massive particle $\hat{H}= \frac{\textbf{p}^2}{2m}+V(x)$), the expression for Wiseman's operationally defined velocity reduces to \citep[p.~5]{wiseman2007grounding}

\begin{align}
\label{gradient weak velocity}
    \textbf{v}(\textbf{x},t)&= \lim\limits_{\tau \rightarrow \nonumber 0}\frac{1}{\tau}(\textbf{x}-\langle\hat{x}_w\rangle)\\\nonumber
    &= \lim\limits_{\tau \rightarrow 0} (\textbf{x}-\Re\frac{\bra{\mathbf{x}}\hat{U}(\tau)\mathbf{\hat{x}}\ket{\psi}}{\bra{\mathbf{x}}\hat{U}(\tau)\ket{\psi}})\\\nonumber
    &= \lim\limits_{\tau \rightarrow 0}\frac{1}{\tau}(\Re\frac{\bra{\mathbf{x}}\mathbf{\hat{x}}\hat{U}(\tau)\ket{\psi}-\bra{\mathbf{x}}\hat{U}(\tau)\mathbf{\hat{x}}\ket{\psi}}{\bra{\mathbf{x}}\hat{U}(\tau)\ket{\psi}})\\\nonumber
    &= \lim\limits_{\tau \rightarrow 0}\frac{1}{\tau}(\Re\frac{\bra{\mathbf{x}}[\mathbf{\hat{x}},\hat{U}(\tau)]\ket{\psi}}{\bra{\mathbf{x}}\hat{U}(\tau)\ket{\psi}})\\\nonumber
    &= \lim\limits_{\tau \rightarrow 0}\frac{1}{\tau}(\Re\frac{\bra{\mathbf{x}}[\mathbf{\hat{x}},\mathds{1}-\frac{i}{\hbar}\hat{H}\tau + \mathcal{O}(\tau^2)]\ket{\psi}}{\bra{\mathbf{x}}\mathds{1}-\frac{i}{\hbar}\hat{H}\tau + \mathcal{O}(\tau^2)\ket{\psi}})\\\nonumber
    &= \Re\frac{\bra{\mathbf{x}}[\mathbf{\hat{x}},-\frac{i}{\hbar}\frac{\hat{p}^2}{2m}]\ket{\psi}}{\psi(x)}\\\nonumber
    &= \Re\frac{\bra{\mathbf{x}}\frac{\hat{p}}{m}\ket{\psi}}{\psi(x)}\\ 
    &=\frac{\hbar}{m}\Im\frac{\nabla \psi(x)}{\psi(x)}=\frac{\hbar}{m}\nabla S(x),
\end{align} where $\nabla S(x)$ is the gradient of the phase of the wave function $\psi(x)$.

\bibliography{library}

\end{document}